\documentclass[superscriptaddress,nofootinbib,amsmath,amssymb,aps,pra, twocolumn,article,longbibliography,floatfix]{revtex4-2}

\usepackage{graphicx}
\usepackage{dcolumn}
\usepackage{bm}
\usepackage{hyperref}
\usepackage[mathlines]{lineno}

\newcommand*\colvec[1]{
        \global\colveccount#1
        \begin{pmatrix}
        \colvecnext
} 
\def\colvecnext#1{
        #1
        \global\advance\colveccount-1
        \ifnum\colveccount>0
                \\
                \expandafter\colvecnext
        \else
                \end{pmatrix}
        \fi
}

\newcommand{\I}{\ensuremath{\mathrm{i}}}

\graphicspath{{./plots/}}

\hypersetup{colorlinks,linkcolor=black,citecolor=blue,urlcolor=blue}

\begin{document}

\title{Fast Quantum Interference of a Nanoparticle via Optical Potential Control}

\author{Lukas Neumeier}
\affiliation{University of Vienna, Faculty of Physics, Vienna Center for Quantum Science and Technology (VCQ), A-1090 Vienna, Austria}
\author{Mario A. Ciampini}
\affiliation{University of Vienna, Faculty of Physics, Vienna Center for Quantum Science and Technology (VCQ), A-1090 Vienna, Austria}
\author{Oriol Romero-Isart}
\affiliation{Institute for Quantum Optics and Quantum Information (IQOQI) Innsbruck, Austrian Academy of Sciences, A-6020 Innsbruck, Austria}
\affiliation{Institute for Theoretical Physics, University of Innsbruck, A-6020 Innsbruck, Austria}
\author{Markus Aspelmeyer}
\affiliation{University of Vienna, Faculty of Physics, Vienna Center for Quantum Science and Technology (VCQ), A-1090 Vienna, Austria}
\affiliation{Institute for Quantum Optics and Quantum Information (IQOQI) Vienna, Austrian Academy of Sciences, A-1090 Vienna, Austria}
\author{Nikolai Kiesel}
\email{nikolai.kiesel@univie.ac.at}
\affiliation{University of Vienna, Faculty of Physics, Vienna Center for Quantum Science and Technology (VCQ), A-1090 Vienna, Austria}

\date{\today}

\begin{abstract}

We introduce and theoretically analyze a scheme to prepare and detect non-Gaussian quantum states of an optically levitated particle via the interaction with a light pulse that generates cubic and inverted potentials. We show that this allows to operate on short time- and lengthscales, which significantly reduces the demands on decoherence rates in such experiments.  Specifically, our scheme predicts the observation of interference of nanoparticles with a mass above $10^8$ atomic mass units delocalised over several nanometers, on timescales of milliseconds, when operated at vacuum levels around $10^{-10}$~mbar and at room temperature. We discuss the prospect of using this approach for coherently splitting the wavepacket of massive dielectric objects using neither projective measurements nor an internal level structure.

\end{abstract}

\maketitle

Optical levitation in high vacuum has become a promising platform for quantum controlling motional states of massive, solid objects \cite{Millen2020,Levitodynamics}. Only recently, different cooling methods enabled the preparation of the quantum ground state of motion of dielectric nanoparticles in the mass range of $10^8$ atomic mass units (a.m.u.) \cite{Delic2020,Magrini2021,Tebbenjohanns2021,Ranfagni2021}. In parallel, spatiotemporal control of optical traps has been used to realize fast changing of complex potential landscapes for the nanoparticle motion \cite{Ciampini2022} -- thereby expanding the available toolbox of optical levitation into the regime of nonlinear dynamics \cite{Rondin2017}. Here we combine these two developments into a new experimental protocol to generate non-Gaussian center-of-mass superposition states at previously unattainable mass and time scales, hence offering access to the high-mass regime of macroscopic quantum phenomena \cite{Leggett2002a,Arndt2014} only by controlling optical potentials.  

\begin{figure*}[t]
    \centering
    \includegraphics[width=2\columnwidth]{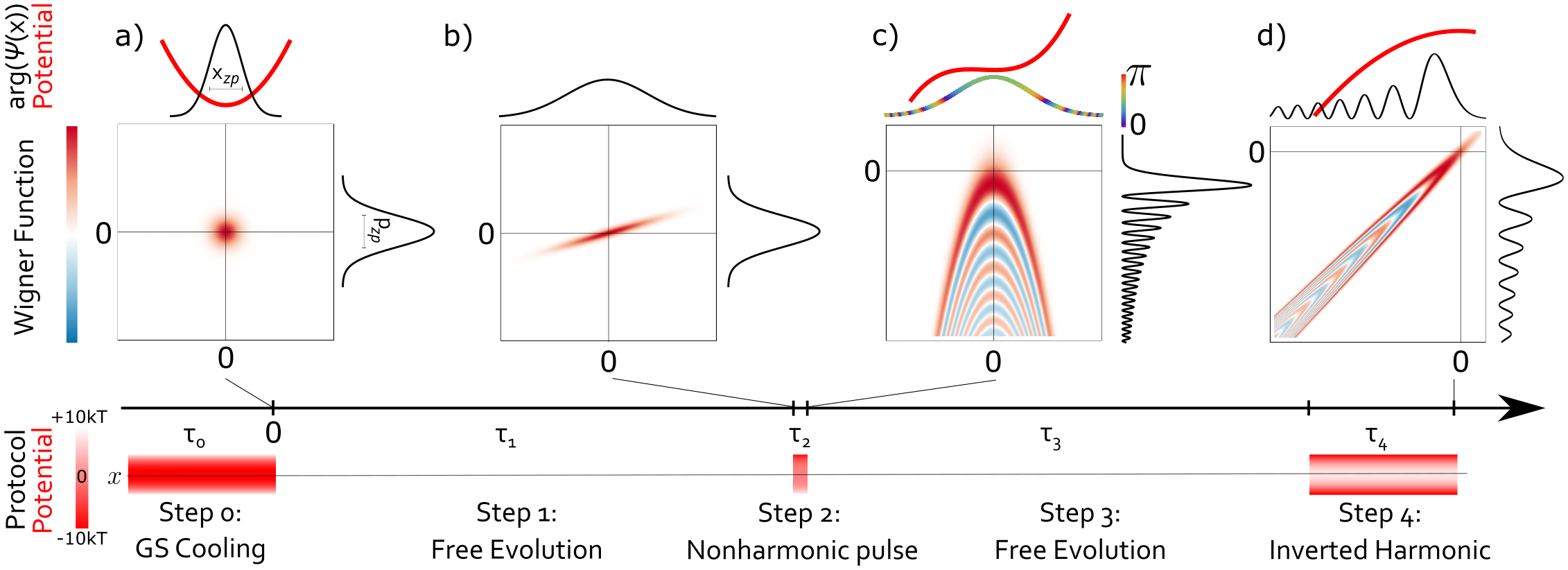}
    \caption{\textbf{Preparation and imaging of a non-Gaussian state.} A wavepacket is initially prepared in the ground state of an harmonic potential (a). After a free expansion of the wavepacket of time $\tau_1$ (b) the particle briefly experiences a cubic and quadratic potential for time $\tau_2$ modulating the phase of the wavepacket while minimizing its momentum uncertainty resulting in a non-Gaussian Wigner function (c). We then image the corresponding momentum space fringe pattern into position space by a free evolution of time $\tau_3$. Finally, we enlarge the fringe spacing to a detectable size with an inverted potential (d).}
    \label{fig1}
\end{figure*}

It is well known that dynamics in non-harmonic potentials allow for quantum state preparation beyond the Gaussian realm \cite{Andrews1997,Schumm2005,Kaufman2015, Radimcubic, Brown2022}. One may be skeptical, however, if the nonlinear features of diffraction-limited optical potentials can be exploited to generate non-Gaussian quantum states given that nanoparticle wavepackets are notoriously small (compared to the optical wavelength) and subject to decoherence from light scattering. In this work we argue that the answer is positive. We overcome both possible showstoppers by using only short interaction times with nonlinear optical potentials and by letting the system freely evolve in-between, which minimizes decoherence by scattering and increases the wavepacket size. In essence, the pulsed light-matter interaction creates a phase modulation of the particle wavepacket \cite{Juffmann2022}. For a cubic potential (and potentials of higher order) this is equivalent to interacting with a sub-wavelength diffraction grating that prepares the desired delocalized non-Gaussian quantum state. This  can be probed in a subsequent measurement through observing a spatial interference pattern. We predict specifically that for a silica particle of radius $r=50$~nm (mass of $\sim 6 \cdot 10^8$~a.m.u.) interference is observable within a few milliseconds. Since operating on such a short time scale dramatically reduces the requirements on decoherence due to gas pressure and blackbody radiation \cite{Weiss2021}, our scheme is predicted to work in a room temperature environment and ultra-high vacuum level ($10^{-10}$~mbar). In comparison to other approaches, our scheme neither relies on external non-linearities, e.g. by coupling to external qubits \cite{GHz,Chu2017,Gieseler2020} or induced by measurement \cite{Romero-Isart2011,Pino2016,Hong2017}, nor on internal degrees of freedom \cite{NV_Li16,Hetet2017, Albrecht14,Urso,Kim2016,Marshman2022}. It only exploits the evolution in a fast, time-varying potential landscape.

The protocol consists of five steps (see Fig.~\ref{fig1}): 0)~Initialization of the ground state in a harmonic potential, 1)~free evolution, 2)~pulsed interaction with a cubic potential, 3)~free evolution, 4)~evolution in an inverted potential. Finally, by switching back to the potential in step 0, the particle position is measured before initializing the state again. Note that in this protocol the particle practically stays at the same place and the process can be repeated immediately and with the same particle. In the following discussion, we do not presume any specific physical implementation until we discuss an optical setting later in the manuscript.

At the heart of the protocol is the preparation of a non-Gaussian state in step 2. The pulsed dynamics in a cubic potential $V_2(x) \propto x^3$ imprints a cubic phase on the particle wavefunction $\psi(x) \rightarrow \psi(x)\exp[-(\I/\hbar) V_2(x) t]$ (see \cite{Radimcubic} for alternative ideas to prepare non-Gaussian states using a cubic potential). In analogy to diffraction at a phase grating \cite{Bateman2014,Fein2019,Brand2020}, the cubic phase results in fringes in momentum space, clearly reflecting its non-Gaussianity (Fig.~\ref{fig1}c). For the fringes to form requires the preparation of a state with sufficient purity, large extension in position and small momentum uncertainty. This is achieved via ground-state cooling in step 0 (Fig.~\ref{fig1}a), a free evolution during step 1 (Fig.~\ref{fig1}b) and introducing an additional quadratic potential during step 2. The fringe pattern can then be observed in position space after another free evolution (step 3) and subsequent expansion of the state (step 4, Fig.~\ref{fig1}d). 

We proceed with a more detailed description of each step. We consider the center-of-mass motion of a particle of mass $m$ along one axis, say $x$. The protocol requires five piece-wise constant potentials $V_i (\hat x)$. The corresponding Hamiltonians are given by $\hat H_i = \hat p^2/(2m) + V_i(\hat x)$, where $\hat x$ and $\hat p$ are the center-of-mass position and momentum operators with $[\hat x, \hat p] = \I \hbar$. 
In step 0 the potential is harmonic $V_0(x) = m \omega_0^2  x^2/2$ with trap frequency $\omega_0$.
In this step, during a time duration $\tau_0$, the particle is prepared in a thermal state with mean phonon occupation $\bar n < 1$ and zero-point motion $x_\text{zp} \equiv [\hbar/(2m \omega_0)]^{1/2}$. In step 1, for a time duration $\tau_1 \gg \omega_0^{-1}$, the potential is switched off, $V_1(x)=0$, to let the position uncertainty $\sigma_x(\tau_1) \approx x_\mathrm{zp}\sqrt{2\bar n+1}\omega_0 \tau_1$ expand linearly in time, while the momentum uncertainty remains constant (Fig.~\ref{fig1}b). In step 2, for a short time duration $\tau_2$, the particle experiences the potential $V_2(x) = m \omega_2^2 x^2/2 + m \omega_2^2 x^3 / l$, where $\omega_2$ defines the stiffness of the harmonic potential and $\sigma_x/l$ quantifies the effect of the cubic potential. This is the key step in the protocol: the phase modulation of the cubic term creates sharp non-Gaussian features in momentum space (Fig.~\ref{fig1}c) if the harmonic term sufficiently reduces the momentum uncertainty by cancelling the quadratic phase acquired during step 1 ($\propto \tau_1^{-1}$).
In step 3, for a time $\tau_3$, the potential is switched off, which maps momentum features into position space while linearly increasing their size. Note that the harmonic term in step 2 also allows to compensate for the quadratic phase ($\propto \tau_3^{-1}$) acquired during step 3. Altogether, after this step the momentum features will be completely mapped into position space if $\omega_2^2 \tau_2 \approx \tau_1^{-1} +  \tau_3^{-1}$ \cite{SM}. In other words, we then expect a sharp image of the interference fringes. In step 4, for a time duration $\tau_4$, we apply an inverted harmonic potential $V_4(x) = - m \omega_4^2 x^2/2$. It expands the position fringes exponentially fast \cite{Pino2016,Romero_Isart_2017} (see Fig.~\ref{fig1}d), and allows to increase the fringe spacing beyond the achievable detection position resolution.
\begin{figure}[b]
    \centering
    \includegraphics[width=0.85\columnwidth]{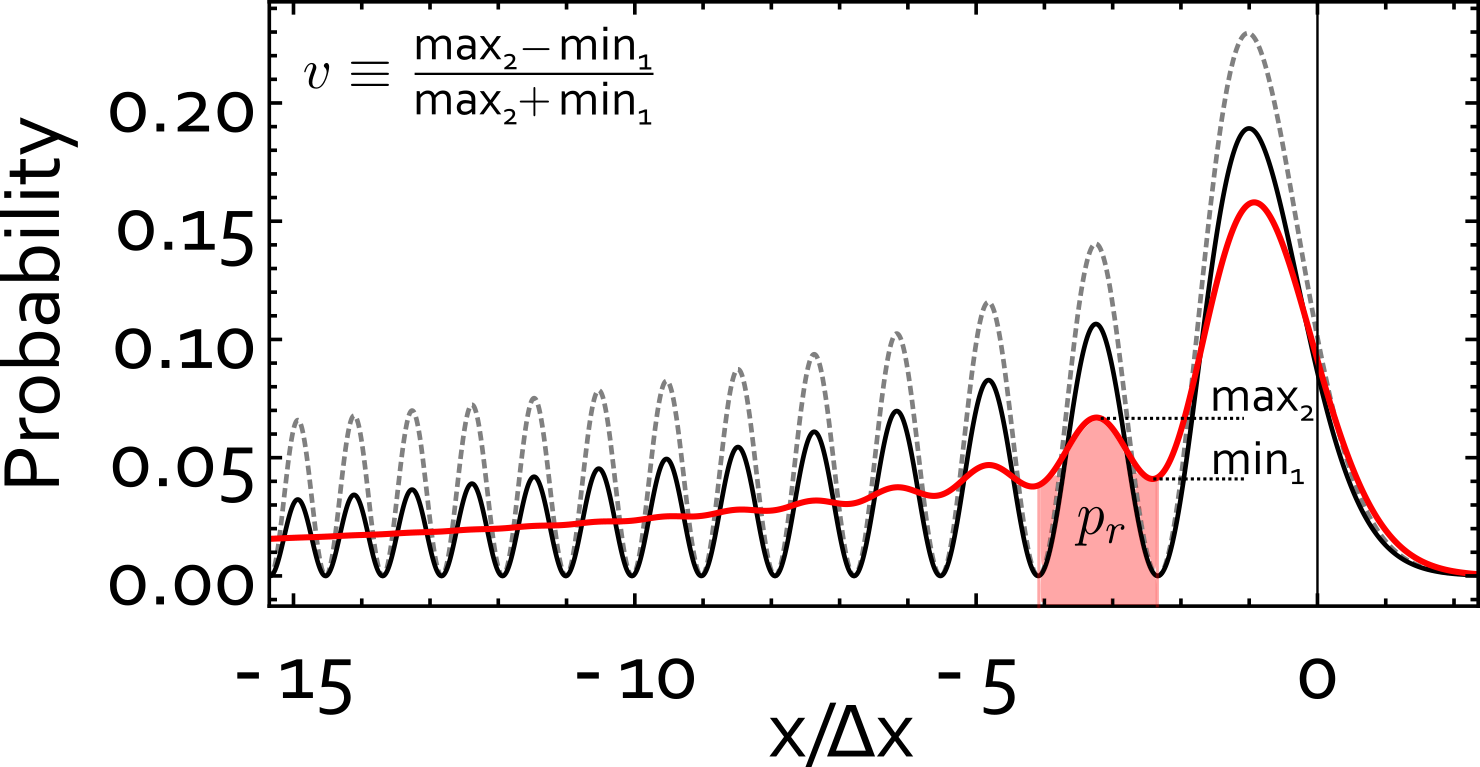}
    \caption{\textbf{Interference Pattern.} Shown is the probability distribution obtained at the end of a complete protocol run with (red) and without (black) decoherence, as well as the underlying square of the Airy-function (dashed-grey). The plot is based on the parameters of our case study (Tab.~\ref{tab}). We use the visibility $v$  and probability to observe the particle in the second fringe $p_r$ (shaded area) to define the valid fringe patterns in Fig.~\ref{fig3} (see \cite{SM}, section VII.)}
    \label{fig2}
\end{figure}

Assuming unitary evolution and a pure initial state, the position probability distribution at the end of the protocol (after step 4) is given by \cite{SM}:
\begin{equation} \label{eq:PxUnitary}
 P(x) = \frac{1}{\sqrt{2 \pi} \Delta x^2 \sigma_c} \left[\mathrm{Ai}\left(\frac{x}{\Delta x} \right) \ast 
 \exp \left( - \frac{x^2}{4 \sigma^2_c} \right)\right]^2,
\end{equation}
where $\ast$ denotes a convolution (with respect to $x$) of an Airy function $\mathrm{Ai}\left(x \right)$ ($\Delta x$ determines the fringe spacing) with a Gaussian of width $\sigma_c$. 
The detailed analytical derivation of Eq.~(\ref{eq:PxUnitary}) is given in \cite{SM} including analytical expressions of the  lengthscales $\Delta x$ and $\sigma_c$. 

Fig.~\ref{fig2} illustrates the probability distribution~(\ref{eq:PxUnitary})~(black line) in comparison to the square of the pure Airy function (dashed grey line), which corresponds to the limit $\Delta x \gg \sigma_c$. Since here we coherently superimpose probability \textit{amplitudes}, the visibility stays perfect even for finite $\sigma_c$, yet the fringe height is exponentially reduced for increasing $|x|$. To observe multiple fringes we require $\Delta x/\sigma_c > 1$, which ensures that the fringe height is not reduced too much. One can determine this ratio in dependence on the protocol parameters as \cite{SM}:
\begin{equation} \label{eq2}
 \frac{\Delta x}{\sigma_c} =  2 \sigma_{x}(\tau_1) \left[  \frac{3 m \omega_2^2 \tau_2}{l \hbar } \right]^{1/3},
\end{equation}
with the position uncertainty $\sigma_{x}(\tau_1)$ after step 1  and a factor that incorporates the interaction strength with the cubic pulse in step 2. Since the latter is mainly limited by the added noise of the actual physical implementation, the condition $\Delta x/\sigma_c > 1$ results in a condition on $\sigma_{x}(\tau_1)$.
 
A major limitation to the free expansion times is decoherence due to gas collisions, as a single collision resolves the particle position. To account for this we limit the total protocol time $\tau_f \equiv \sum^4_\mathrm{i=1} \tau_i $ such that no gas collision occurs in $90\%$ of the protocol runs (\cite{SM}, section V) and omit this source of decoherence in the further discussion. A second, universal source of decoherence is scattering, emission and absorption of black-body radiation \cite{ORIPRA11}. It can be described with a master equation of the form 
\begin{equation} \label{eqn:dec}
\dot{\rho} = - \frac{\mathrm{i}}{\hbar} [\hat H_i ,\hat \rho] -\Lambda_i [\hat x,[\hat x, \hat \rho]].    
\end{equation}
Here $\hat \rho$ is the density matrix of the center-of-mass state, and $\Lambda_i$ the localization rate. Both are assumed to be constant during each step $i$ of the protocol. Depending on the actual implementation the localization rates may also include other sources of decoherence (e.g. recoil heating).

\begin{table*}
\centering
\begin{tabular}{|cccccc|c|ccc|c|cc|ccc|} 
\hline
\multicolumn{6}{|c|}{Initial Conditions}                                         & \multicolumn{7}{c}{Protocol Parameters}                                                                         & \multicolumn{3}{|c|}{Results}  \\ 
\hline
$T_e [K]$ & $P [\mathrm{mbar}]$ & $\lambda [\mathrm{nm}]$ & $r [\mathrm{nm}]$ & $\bar{n}$ & $\omega_0 [\mathrm{kHz}]$ & $\tau_1 [\mathrm{ms}]$ & $\omega_2 [\mathrm{kHz}]$ & $\phi_2$  & $\tau_2 [\mathrm{\mu s}]$ & $\tau_3 [\mathrm{ms}]$ & $\omega_4 [\mathrm{kHz}]$ & $\tau_4 [\mathrm{ms}]$ & $T_i [K]$ & $p_r$ & $v$       \\ 
\hline
$300$       & $10^{-10}$ & $1550$          & $50$       & $0.5$       & $2\pi \cdot 100$ & $1.34$          & $2\pi \cdot 2.5$ & $0.05\pi$ & $10$            & $0.66$          & $2\pi \cdot 10$  & $0.087$         & $315.2$     & $0.094$ & $0.23$      \\
\hline
\end{tabular}
\caption{
Parameter set for the case study. The interference pattern expected for this parameter set is shown in Fig.~2, red line. Properties of silica nano-particles: density $\rho = 1850 \,~\mathrm{kg/m^3}$, heat capacity: $c_v = 700$~J/K, refractive index at $\lambda = 1550$~nm: $n_\mathrm{1550} = 1.43 + \I 2.46 \cdot 10^{-9}$. }
\label{tab}
\end{table*}

The effect of decoherence as described by eq.~(\ref{eqn:dec}) and assuming an initially mixed Gaussian state can be calculated analytically. For $\Lambda_1 x_\mathrm{zp}^2 \tau_1 \ll 1$ , the final position probability distribution is given by (\cite{SM}, section II):
\begin{equation} \label{eq:PxDecoh}
P_D(x) =  \frac{P(x)}{\sqrt{2 \pi \sigma_\Lambda^2}} \ast \exp \left( - \frac{x^2}{2\sigma_\Lambda^2} \right),
\end{equation}
which is a convolution of the position probability distribution obtained by the unitary dynamics in~eq.~(\ref{eq:PxUnitary}) with a Gaussian of variance $\sigma_\Lambda^2$ reducing the visibility of the interference fringes (as shown in Fig.~\ref{fig2}, red line). The key parameter $\sigma_\Lambda$ depends on the localisation rates during each of the protocol steps and the phonon occupation number $\bar{n}$ of the initial thermal state. Most importantly, for a finite visibility $v>0$ we require $\Delta x > \sigma_\Lambda$ (Fig.~S5 in \cite{SM}). 

\begin{figure}[t]
    \centering
    \includegraphics[width=0.85\columnwidth]{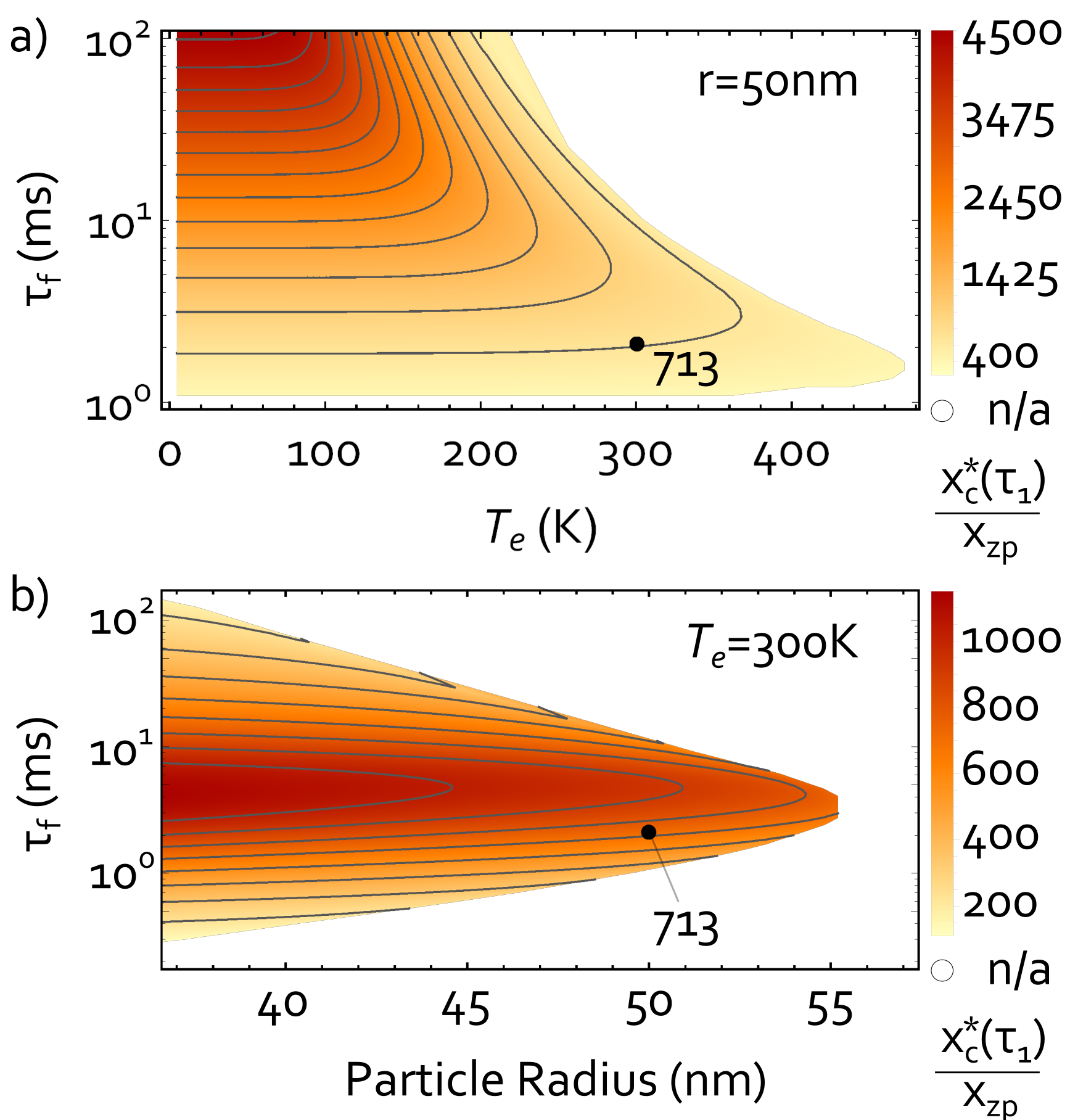}
    \caption{\textbf{Certified coherence length:} We show the lower bound on coherence length $x_c^*(\tau_1)$ that we can certify with our scheme in dependence on a) the environmental temperature $T_e$ for a particle of $r=50$~nm and b) the particle radius for $T_e=300$~K. The plots include the influence of decoherence from black-body radiation and photon recoil. The vertical axis shows the total protocol time $\tau_f$ for a single protocol run, respectively. We fix $v^2 p_r = 0.005$ and a fringe spacing of $5$~nm for all data-points (Details on optimization in~\cite{SM}, section VIII). 
    \label{fig3}}
\end{figure}


For a further analysis we need to be more specific about the physical implementation. Here we consider a dielectric nanoparticle in a time-dependent potential that is realized with an optical standing wave: $V_i(x) = - (m \omega_i^2 \lambda^2)/(8\pi^2)\cos^2(2 \pi x /\lambda+ \phi_i)$. 
The potentials $V_i(x)$ for each step can be obtained by controlling the laser intensity (proportional to $\omega_i^2$) and the phase $\phi_i$ of the standing wave (\cite{SM}, section V) generating the harmonic (step 0: $\phi_0=0$), the harmonic + cubic (step 2: $0<\phi_2< \pi/8$) and the inverted (step 4: $\phi_4=\pi/2$) potential. We additionally require electronic control of the charged particle for three purposes: feedback-based ground state cooling \cite{Magrini2021, Tebbenjohanns2021}, compensation of the linear part of the optical potential during step 2 (to keep the particle inside the trapping volume), and stabilization against gravity in the vertical direction. 

As laser light exposure increases the internal temperature $T_i$ of the nanoparticle by absorption, $\Lambda_1$ and $\Lambda_3$ are dominated by black-body radiation (\cite{SM}, section III), while $\Lambda_2$ and $\Lambda_4$ are dominated by scattering of laser photons (\cite{SM}, section IV). For position detection, the optical field is switched back to step 0 and the first time-period of detection $t_D \ll 2\pi/\omega_0$ can be used for position readout. Finite detection resolution will be compensated by the exponential increase of the size of the interference pattern in step 4. We design the protocol to confirm the existence of interference fringes with a $5 \sigma$~confidence after approximately $1.2\cdot 10^4$ experimental runs which is achieved for $v^2 p_r = 0.005$ corresponding to less than 1 minute total measurement time (\cite{SM}, section VII/VIII; $p_r$ is defined in Fig.~\ref{fig2}).

\begin{figure}[t]
    \centering
    \includegraphics[width=0.85\columnwidth]{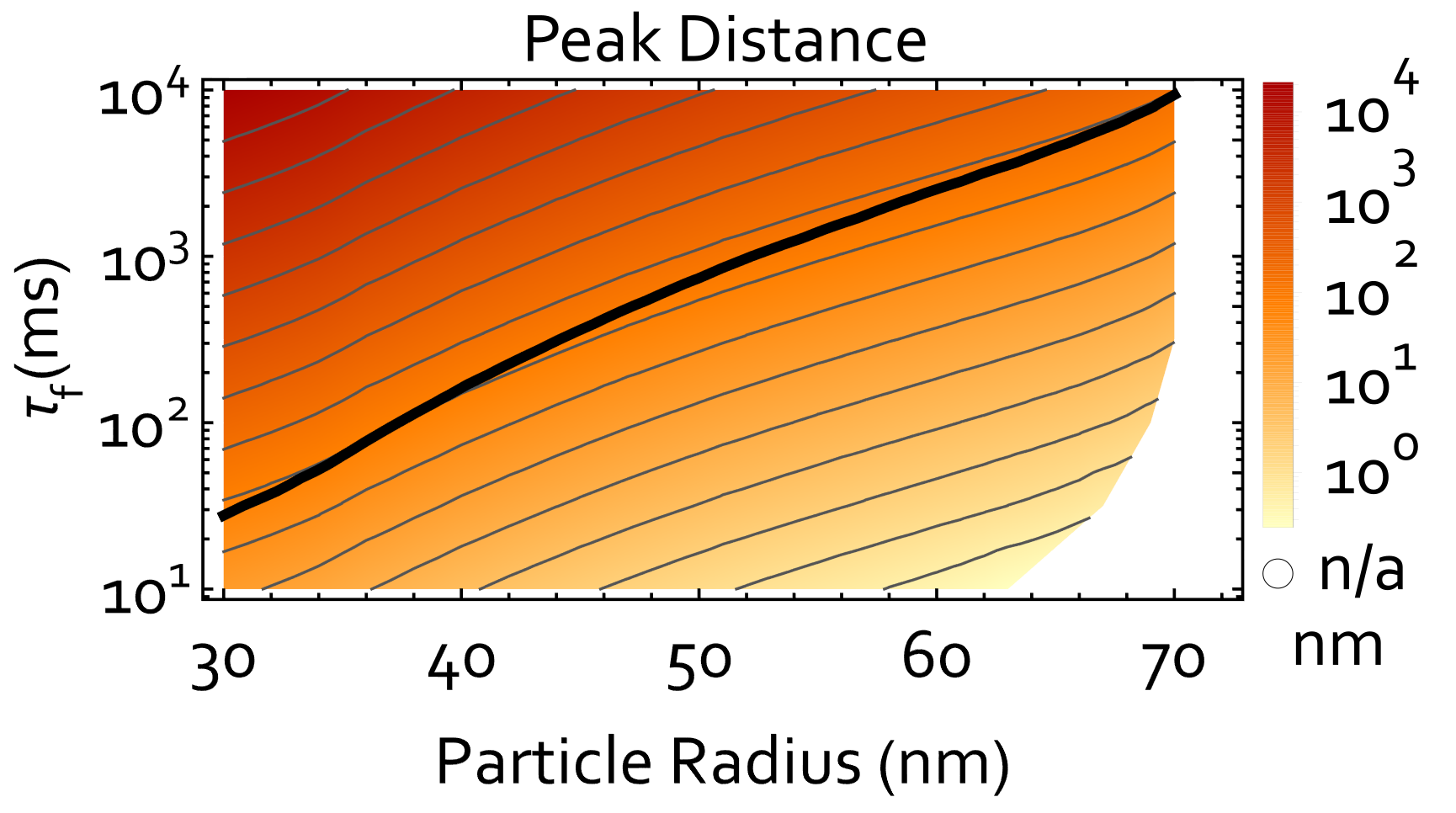}
    \caption{\textbf{Coherent splitting:} We show the distance between the first two intensity maxima ($x_\mathrm{max1}$, $x_\mathrm{max2}$) of the interference patterns for protocols that preserve coherence between these maxima ($g_1(x_\mathrm{max2},x_\mathrm{max1}) > 0.95$) for different particle radii $r$ and protocol times $\tau_f$. Here, we do not include decoherence due to black-body radiation and omit step 4 (inverted potential). The black line shows the parameter set where the coherent splitting between the fringes corresponds to the particle diameter.
    \label{fig4}}
\end{figure}

Creating quantum interference during step 2 requires the wave packet after step 1 to exhibit sufficient spatial coherence, where we quantify the coherence length $x_c$ as the width of the correlation function $g_1(x_2 - x_1)=\exp(-(x_2 -x_1)^2/(2 x_c^2))$, in full analogy to optical coherence \cite{Glauber} (\cite{SM}, section VI).
In turn, we can certify a lower bound on the coherence length before step 2 $x_c^*(\tau_1) \equiv 2 \sigma_x(\tau_1) \sigma_c/\sigma_\mathrm{\Lambda} < x_c(\tau_1)$ (\cite{SM}, section VI) based on the experimental observation of the final shape of the interference pattern and on the width $\sigma_x(\tau_1)$ of the state after the first free expansion. For example, the observation of the probability density in Fig.~\ref{fig2}~(red line) implies  $x_c^*(\tau_1) \approx 713\, x_\mathrm{zp} \approx 6.6$~nm. 

In Fig.~\ref{fig3}, we study the dependence of $x_c^*(\tau_1)$ on the particle radius $r$, environmental temperature $T_e$ and total protocol time $\tau_f$, effectively determining the pressure required for the experiment. The criteria on the interference correspond to those described for the case study and Fig.~\ref{fig2}. See \cite{SM},~section VIII, for details on the optimization.
In Fig.~\ref{fig3}a we vary the particle radius $r$ at room temperature ($T_e=300$~K), for Fig.~\ref{fig3}b we vary the environmental temperature $T_e$ for a radius of $r=50$~nm. In both panels our case study is marked with a black dot. It can be seen that for our criteria, larger particle sizes are not accessible at room temperature and also superposition sizes are limited. Yet, cryogenic temperatures promise to enable the verification of much larger superpositions with our scheme.

Looking ahead, one may ask if the optical cubic potential might also serve to \textit{coherently} split the wavepacket, ideally even beyond the particle size. More precisely, under which conditions are the two main peaks in the interference pattern Fig.~\ref{fig2} coherent ($g_1(x_\mathrm{max1}, x_\mathrm{max2}) > 0.95$) and separated by more than the particle diameter ($|x_\mathrm{max1}-x_\mathrm{max1}| > 2 r$). To achieve this we need to avoid the optical inverted potential in step 4 ($\tau_4=0$), as it localizes the state. Since without step 4, $\Delta x$ grows only linearly in time (\cite{SM}, section IX), this requires much longer protocol times $\tau_f$. To determine protocols for coherent splitting, we neglect decoherence/scattering rates due to black-body radiation/gas molecules in the calculation and note that they are bound by the inverse protocol time, respectively. We still assume $\bar{n}=0.5$. The result of this analysis is shown in Fig.~\ref{fig4}.  The analysis shows that optical control is compatible with a splitting of the particle wave packet even beyond the particle size (above the black line). The corresponding decoherence rates are very demanding and can likely only be achieved at cryogenic temperatures of the environment and the internal particle temperature.

In summary, we presented a scheme to prepare and observe non-Gaussian features of a massive particle which is based on the control of pulsed potentials. We provided a platform-independent analytical analysis of our protocol. We theoretically showed that, despite of photon recoil, an optical implementation of our protocol can generate matter-wave interference for a silica nanoparticle with a mass exceeding $\sim 10^8$ atomic mass units at room temperature and with feasible vacuum levels. This significant reduction of environmental demands is enabled by operating at short length- and timescales. We investigated how the generation of the interference pattern depends on the particle size and environmental temperature. In addition, we analysed the extent of nanoparticle delocalization that is detectable via the interference pattern and characterized the environmental conditions that would be required to observe coherent splitting of the wavepacket beyond the particle size. We believe that the use of pulsed potentials provides a central new element in the toolbox of levitated optomechanics. An experimental implementation of this proposal would provide a remarkable step in the field of macroscopic quantum physics.

{\bf Acknowledgements}. We thank Thomas Agrenius, Davide Candoli, Darrick Chang, Fulvio Flamini, Florian Marquardt,  Lea Trenkwalder and Klemens Winkler for insightful discussions. This
research was funded in  part by the Austrian Science Fund (FWF START Project TheLO, Y 952-N36) and the European Research Council (ERC Synergy Q-Xtreme). MAC acknowledges support from the FWF Lise Meitner Fellowhip (M2915, ”Entropy generation in nonlinear levitated optomechanics”).

\bibliographystyle{apsrev4-2}
\bibliography{main}

\end{document}


\title{Supplemental Material: Fast Quantum Interference of a Nanoparticle via Optical Potential Control}

\author{Lukas Neumeier}
\affiliation{University of Vienna, Faculty of Physics, Vienna Center for Quantum Science and Technology (VCQ), A-1090 Vienna, Austria}
\author{Mario A. Ciampini}
\affiliation{University of Vienna, Faculty of Physics, Vienna Center for Quantum Science and Technology (VCQ), A-1090 Vienna, Austria}
\author{Oriol Romero-Isart}
\affiliation{Institute for Quantum Optics and Quantum Information (IQOQI) Innsbruck, Austrian Academy of Sciences, A-6020 Innsbruck, Austria}
\affiliation{Institute for Theoretical Physics, University of Innsbruck, A-6020 Innsbruck, Austria}
\author{Markus Aspelmeyer}
\affiliation{University of Vienna, Faculty of Physics, Vienna Center for Quantum Science and Technology (VCQ), A-1090 Vienna, Austria}
\affiliation{Institute for Quantum Optics and Quantum Information (IQOQI) Vienna, Austrian Academy of Sciences, A-1090 Vienna, Austria}
\author{Nikolai Kiesel}
\affiliation{University of Vienna, Faculty of Physics, Vienna Center for Quantum Science and Technology (VCQ), A-1090 Vienna, Austria}

\date{\today}

\maketitle

\section{Protocol Without Decoherence}
Here we derive Eq.~(1) in the main text. As initial state (step 0) we assume the ground state of an harmonic oscillator with mechanical frequency $\omega_0$ and unitary evolution in step 1,2,3 and 4. In this derivation we omit normalization factors and normalize the end result.
\subsubsection{Step 0: Groundstate Cooling}
We begin with the groundstate of an harmonic oscillator
\begin{equation}
    \Psi_0(x) \propto \exp \left(-\frac{x^2}{4x_\mathrm{zp}^2} \right), 
\end{equation}
with zero-point motion $x_\text{zp} \equiv [\hbar/(2m \omega_0)]^{1/2}$.
\subsubsection{Step 1: Free Evolution for $\tau_1$}
Using the free-evolution propagator we obtain:

\begin{equation}
    \Psi_1(x) \propto \exp \left( - \frac{x^2}{4 \sigma_x^2(\tau_1)} + \I \frac{\tau_1 \omega_0}{4 \sigma_x^2(\tau_1)} x^2  \right),  
\end{equation}
with position variance $\sigma_x^2(\tau_1) = x_\mathrm{zp}^2(1 + \omega_0^2 \tau_1^2)$.
\subsubsection{Step 2: Cubic + Harmonic Pulse for $\tau_2$}
Since we assume this pulse to be sufficiently short (for requirements, see \ref{short}), we neglect the kinetic part of the Hamiltonian
$H_2 \approx V_2(x) = m \omega_2^2 x^2/2 + m \omega_2^2 x^3/l $
which results in a multiplication with a quadratic + cubic phase:
\begin{equation}
    \Psi_2(x) =  \exp \left(- \frac{\I}{2 \hbar }m \omega_2^2 x^2 \tau_2 -  \frac{\I}{\hbar l }m \omega_2^2  x^3 \tau_2 \right) \Psi_1(x).
\end{equation}

\subsubsection{Step 3 and 4: Free-Evolution for $\tau_3$ followed by an inverted harmonic potential for $\tau_4$}
We again use the propagator for free-evolution (without solving the convolution-integral) and then the propagator for the evolution in an harmonic potential \cite{condon1937immersion} while substituting the real mechanical frequency with an imaginary mechanical frequency $\I \omega_4$, which describes the evolution in an inverted harmonic potential. We solve the second integral first since it only involves Gaussians, obtaining 
\begin{equation}\label{eq1}
    \Psi_4(x) \propto \exp\left(\I \frac{m \omega_4}{2 \hbar} x^2 \right) \int dx_1 \exp \left( -  \frac{\I}{\hbar l }m \omega_2^2  x_1^3 \tau_2  - \frac{x_1^2}{4 \sigma_x^2(\tau_1)} + \I b_c x_1^2 + \I \frac{2}{\hbar} \frac{m\omega_4\exp(-\omega_4 \tau_4)}{1 +  \omega_4 \tau_3 } x x_1\right),
\end{equation}
where we assumed $\omega_4 \tau_4 \gg 1$ and with 
\begin{equation}
    b_c = \frac{m}{2 \hbar}\left(\frac{\omega_4}{1+\tau_3 \omega_4} + \frac{\tau_1 \omega_0^2}{\sigma_x^2(\tau_1)}-\omega_2^2 \tau_2\right).
\end{equation}
Using the convolution theorem in Eq.~(\ref{eq1}), yields
\begin{equation}\label{psi4}
 \Psi(x,\tau_4)  \propto \exp \left(\mathrm{i} \frac{ m \omega_4 x^2 }{2 \hbar}\right) \left[\mathrm{Ai}\left(\frac{x}{\Delta x} \right) \ast \exp \left(- \frac{x^2}{4 \sigma^2_c} \right) \ast \exp \left(\I \frac{x^2}{4 \sigma^2_\mathrm{cb}} \right) \right]. 
\end{equation}
 Here $\ast$ denotes a convolution with respect to $x$ and $\mathrm{Ai}\left(\frac{x}{\Delta x} \right)$ is the Airy-function,
with
\begin{equation}\label{deltax}
 \frac{\Delta x}{\sigma_c} = 2 \sigma_x(\tau_1) \left[\frac{3 m  \omega_2^2 \tau_2}{\hbar l} \right]^{1/3},
\end{equation} 
\begin{equation}\label{ss}
\sigma_c = \frac{\hbar}{4 \sigma_x(\tau_1)} \frac{\omega_4 \tau_3 + 1}{m \omega_4} e^{ \omega_4 \tau_4},
\end{equation}
and 
\begin{equation}
    \frac{\sigma_\mathrm{bc}}{\Delta x} = \sqrt{\frac{b_c}{4}} \left[\frac{3 m  \omega_2^2 \tau_2}{\hbar l} \right]^{-1/3}.
\end{equation}
As the convolution with the quadratic phase reduces the visibility of interference fringes by lifting them off zero, we require $\sigma_\mathrm{bc}/\Delta x < 1$. Then this convolution can be neglected as a convolution with oscillations on length scales much smaller than $\Delta x$ average themselves out and the quadratic phase can be approximated by a delta function. This condition can be easily fullfilled by setting $b_c \approx 0$, which requires choosing
\begin{equation}\label{cond}
    \omega_2^2 \tau_2 = \frac{\omega_4}{1 + \omega_4 \tau_3} + \frac{\omega_0^2 \tau_1}{\sigma_x^2(\tau_1)} \approx \frac{1}{\tau_3} + \frac{1}{\tau_1},
\end{equation}
where we assume $\omega_0 \tau_1 \gg1$ and $\omega_4 \tau_3 \gg 1$.
Normalizing for $b_c \approx 0$ yields:
\begin{equation}\label{psi41}
 \Psi(x,\tau_4) = \frac{1}{(2 \pi)^{1/4} \Delta x \sqrt{\sigma_c} } \exp \left(\mathrm{i} \frac{ m \omega_4 x^2 }{2 \hbar}\right) \left[\mathrm{Ai}\left(\frac{x}{\Delta x} \right) \ast \exp \left(- \frac{x^2}{4 \sigma^2_c} \right) \right], 
\end{equation}
which absolute value square produces Eq.~(1) in the main text.
For completeness we provide the first two moments in position and momentum:
\begin{equation}
    \langle x(\tau_4) \rangle = -\frac{\Delta x^3}{4 \sigma_c^2},
\end{equation}
and
\begin{equation}
    \langle x^2(\tau_4) \rangle =
    \sigma_c^2 \left(1+ \frac{3\Delta x^6}{16 \sigma_c^6} \right),
\end{equation}
\begin{equation}
    \langle p(\tau_4) \rangle = -\frac{m \omega_4 \Delta x^3}{4 \sigma_c^2},  
\end{equation}
\begin{equation}
    \langle p^2(\tau_4) \rangle = \frac{\hbar^2}{4 \sigma_c^2}   + \frac{3 m^2 \Delta x^6 \omega_4^2}{16 \sigma_c^4}+ m^2 \omega_4^2 \sigma_c^2. 
\end{equation}
\section{Position probability distribution for an initial thermal state subject to displacement noise}
Here we derive Eq.~(4) of the main text assuming as master equation of the form of Eq.~(3) in the main text for each step and an initial Gaussian (e.g. thermal) state. Note that decoherence due to laser photons, black-body radiation, electric/magnetic field fluctuations and various linear white noise sources like linear shot/intensity noise can all be modeled by such a master equation (displacement noise) as long as the interacting particles cannot resolve the position of the particle better than its standard deviation.  
We separate the derivation into two parts: 
\begin{itemize}
    \item Before the non-linear interaction (before step 2), where the state is Gaussian and the effect of decoherence can be quantified by its size $\sigma_x(\tau_1)$ and the decay of its purity $P(\tau_1) = \mathrm{Tr}(\rho^2(\tau_1))$.
    \item During and after the non-linear interaction (during and after step 2), where $\sigma_x(\tau_1)$, $P(\tau_1)$ and decoherence/noise during step 2, 3 and 4 can be absorbed into a single total blurring distance $\sigma_\Lambda$, which reduces the visibility of the final interference peaks. 
\end{itemize}
\subsubsection{Initial thermal state (step 0) and displacement noise during the first free evolution (step 1)}
The time evolution of the Wigner function describing any Gaussian state evolving with a quadratic Hamiltonian can be expressed as
\begin{equation}
 W(x,p,t) = \frac{\sqrt{4 a_1(t) a_2(t) - a_3^2(t)}}{2 \pi} \exp{\left(- a_1(t) x^2 - a_2(t) p^2 - a_3(t) p x\right)},
\end{equation}
with
\begin{equation}
a_1(t) = \frac{2 P^2(t)\sigma_p^2(t)}{\hbar^2},
\end{equation}
\begin{equation}\label{genwig}
a_2(t) = \frac{2 P^2(t) \sigma_x^2(t)}{\hbar^2}
\end{equation}
and
\begin{equation}
a_3(t) = \pm \frac{2 P^2(t) \sqrt{ 4 \sigma_x^2(t) \sigma_p^2(t) - \hbar^2/P^2(t)}}{\hbar^2},
\end{equation}
where the sign of $a_3(t)$ is the sign of $\langle xp + px \rangle$, $\sigma_x^2(t)$ is the variance in position space and $\sigma_p^2(t)$ is the variance in momentum space. 
Thus, every Gaussian state is fully characterized by only three parameters (in the reference frame where $\langle x \rangle = \langle p \rangle = 0$).

With $P(0) = (2 \bar{n}+1)^{-1}$, $\sigma_x^2(0) =  x_\mathrm{zp}^2 (2 \bar{n} +1 )$, $\sigma_p^2(0) = p_\mathrm{zp}^2(2 \bar{n} +1 ) $, 
$x_\mathrm{zp}= \sqrt{\hbar/(2 m \omega_0)}$, and $p_\mathrm{zp}= \sqrt{\hbar \omega_0 m/2 }$,
we specify our initial state $W(x,p,0)$ as a thermal harmonic oscillator with mechanical frequency $\omega_0$ and average phonon occupation number $\bar{n}$. Note that for a thermal harmonic oscillator state $a_3(0) = 0$.

We proceed by calculating $P(\tau_1)$, $\sigma_x(\tau_1)$ and $\sigma_p(\tau_1)$ after the first free evolution (step 1) during which the particle experiences displacement noise (typically dominated by the emission of black-body radiation) with localization rate $\Lambda_1$. The time evolution of its Wignerfunction $W = W(x,p,t)$ is governed by \cite{soto1983time}
\begin{equation}
\frac{\partial W}{\partial t} = -\frac{p}{m} \frac{\partial W}{\partial x} + \hbar^2 \Lambda_1 \frac{\partial^2 W}{\partial p^2}.
\end{equation}
To solve this equation we switch the state representation to \cite{Pino2016}
\begin{equation}
\tilde{W}(k_x,k_p,t) = \frac{1}{2 \pi}\int dx dp \exp\left(- \I k_x x - \I k_p p\right) W(x,p,t),
\end{equation}
which evolves with a simpler equation
\begin{equation}\label{frr}
\frac{\partial \tilde{W}}{\partial t} = -\frac{k_x}{m} \frac{\partial \tilde{W}}{\partial k_p} - \hbar^2 \Lambda_1 k_p^2 \tilde{W}.
\end{equation}
We make a separation ansatz
\begin{equation}\label{ansatz}
\tilde{W} = \tilde{W}(k_x,k_p,t) = \tilde{W}_\mathrm{\Lambda_1=0}(k_x,k_p,t) \tilde{G}(k_x,k_p,t),
\end{equation}
where $\tilde{W}_\mathrm{\Lambda_1=0}(k_x,k_p,t)$ describes free evolution without decoherence. 
$\tilde{G}(k_x,k_p,t)$ satisfies the same equation as $\tilde{W}$ with initial condition $G(k_x,k_p,0) = 1$. Solving Eq.~(\ref{frr}) for $G(k_x,k_p,0)$ yields 
\begin{equation}
\tilde{G}(k_x,k_p,t) = \exp\left(- \hbar^2 \Lambda_1 t k_p^2 - \frac{\hbar^2 \Lambda_1 t^2}{m} k_p k_x-\frac{\hbar^2 \Lambda_1 t^3}{3 m^2} k_x^2 \right).
\end{equation}
To calculate the Wigner function after $\tau_1$ we double inverse Fourier transform Eq.~(\ref{ansatz}) back into the Wigner representation
\begin{equation}\label{wig1}
W_1(x,p,\tau_1) = \frac{1}{2 \pi} \int dk_x dk_p e^{\I k_x x + \I k_p p} \tilde{W}_\mathrm{\Lambda_1=0}(k_x,k_p,\tau_1) \tilde{G}(k_x,k_p,\tau_1),
\end{equation}
where $ \tilde{W}_\mathrm{\Lambda_1=0}(k_x,k_p,\tau_1) $ is the double Fourier transform of $W(x - (p/m) \tau_1, p,0)$.

After computing the Gaussian integrals of Eq.~(\ref{wig1}),
we can identify $P^2(\tau_1) \sigma_x^2(\tau_1) = \hbar^2 a_2(\tau_1)/2$ with Eq.~(\ref{genwig}), where $a_2(\tau_1)$ is the coefficient of $p^2$ in the exponent.
For $\omega_0^2 \tau_1^2 \gg 1$ and $x_\mathrm{zp}^2 \Lambda_1 \tau_1 \ll 1$, we obtain
\begin{equation}
P(\tau_1)\approx \left( \frac{8}{3} \Lambda_1 \tau_1  \sigma_x^2(\tau_1) + (2 \bar{n} + 1)^2 \right)^{-\frac{1}{2}}.
\end{equation}
We see how the purity gets reduced by displacement noise during a free evolution, while for $\omega_0^2 \tau_1^2 \gg 1$ and $x_\mathrm{zp}^2 \Lambda_1 \tau_1 \ll 1$
the position uncertainty $\sigma_x(\tau_1)$ increases approximately linearly since
\begin{equation}
    \sigma_x^2(\tau_1) =\int dp dx x^2 W(x,p,\tau_1)= x_\mathrm{zp}^2(2 \bar{n} + 1) (\omega_0^2 \tau_1^2 + 1) + 2 \Lambda_1 \hbar^2 \tau_1^3/(3 m^2) \approx x_\mathrm{zp}^2(2 \bar{n} + 1) (\omega_0^2 \tau_1^2 + 1),
\end{equation}
as does the momentum space variance
\begin{equation}
    \sigma_p^2(\tau_1) =\int dp dx p^2 W(x,p,\tau_1)= p_\mathrm{zp}^2(2 \bar{n} + 1) + 2 \Lambda_1 \hbar^2 \tau_1.
\end{equation}
These equations reproduce results previously obtained by \cite{romero2011quantum}.

\subsubsection{Displacement noise during the cubic + harmonic pulse  (step 2)}
As we now know the Gaussian state after step 1, we take it as our new initial state in the following derivation with corresponding values $a_1 \equiv a_1(\tau_1)$, $a_2 \equiv a_2(\tau_1)$ and $a_3 \equiv a_3(\tau_1)$.
We begin by switching into the B-representation of quantum phase space \cite{cabrera2015efficient} defined as
\begin{equation}
 B(x,\Theta) = \int dp W(x,p) e^{- \I p \Theta},     \end{equation} 
leading to
\begin{equation}
  B_1(x, \Theta) = \exp \left[ - \left(a_1 - \frac{a_3^2}{ 4 a_2} \right) x^2 - \frac{1}{ 4 a_2} \Theta^2 - \I \frac{a_3}{2 a_2} x \Theta \right],
\end{equation}
which evolves under the influence of displacement noise  with localization rate $\Lambda_2$ with
\begin{equation}
 \dot{B}(x,\Theta) = \left[ \frac{- \I}{m} \frac{d^2}{dxd\Theta} + \frac{\tilde{V}}{\I \hbar} - \hbar^2 \Lambda_2 \Theta^2 \right] B(x, \Theta).
\end{equation}
In step 2, $\tilde{V}=V^- - V^+=- 2 \hbar u_2 x \Theta -3 u_3 \hbar x^2 \Theta  - u_3 \hbar^3 \Theta^3/4$ since $V^\pm = V_2(x \pm \hbar \Theta/2)$, where $V_2(x) = u_2 x^2 + u_3 x^3$ describes the experienced potential at step 2.
Aligned with our previous assumption of short pulses, we approximate $(d/dx) B(x,\Theta) = 0$. Thus,
\begin{equation}\label{bba}
 B_2(x,\Theta,\tau_2) = \exp\left[ -\frac{\I}{\hbar} \tilde{V} \tau_2 - \hbar^2 \Lambda_2 \Theta^2 \tau_2 \right]  B_1(x,\Theta),  
\end{equation}  
 and
 \begin{equation}\label{wigner}
  W_2(x,p,\tau_2) = e^{- (a_1 - \frac{a_3^2}{ 4 a_2} ) x^2} \int d \Theta \exp\left[\I \left(p + 3 \I \tau_2 u_3 x^2 - \frac{a_3}{2 a_2} x \right)\Theta - \left(\frac{1}{4 a_2} + \frac{\sigma_2^2}{2} \right) \Theta^2 + \frac{\I}{4} \hbar^2 \tau_2 u_3   \Theta^3 + 2 \I \tau_2 u_2 x \Theta  \right],
 \end{equation}
with 
\begin{equation}\label{sigma2}
    \sigma_2^2= 2 \hbar^2 \Lambda_2 \tau_2, 
\end{equation}
being the momentum blurring variance caused by displacement noise during step 2.

To see how an initial thermal state and decoherence during
step 1 and 2 affects the final position probability distribution, we proceed by substituting the classical solutions of the Hamilton equations describing evolution during step 3 and 4 into the Wigner function.
Thus, in order to account for the second free evolution (step 3), we replace $x$ by $x - (\tau_3/m) p$ and then, in order to continue with the inverted potential in step 4,
 $x$ by $x_0$ with
\begin{equation}
 x_0 =x \cosh{(\omega_4 \tau_4)} - \frac{p}{m \omega_4}\sinh{(\omega_4 \tau_4)},
\end{equation}
and then all
$p$ by $p_0$ with
\begin{equation}
 p_0 =  p \cosh{(\omega_4 \tau_4)} -  m \omega_4 x \sinh{(\omega_4 \tau_4)}.
\end{equation}
Next, we take the coefficient in the exponent which is proportional to $p \Theta$, set it to zero, solve for $u_2$ which is given by
\begin{equation}
u_2 = -\frac{a_3}{4 a_2 \tau_2} + \frac{m \omega_4}{2 \tau_2 \tau_3 \omega_4 + \tau_2 \tanh(\omega_4 \tau_4)},
\end{equation} 
and substitute the remaining $u_2$ with this expression in our current Wigner function Eq.~(\ref{wigner}).
  For $x_\mathrm{zp}^2 \Lambda_1 \tau_1 \ll 1$ and $\omega_0^2 \tau_1^2 \gg 1$ we obtain $-\frac{a_3}{2 a_2 m}\approx 1/\tau_1$, which with
 $u_2 = \omega_2^2 m/2$ and $\omega_4 \tau_4 \gg 1$,
leads to the condition
\begin{equation}
 \omega_2^2 \tau_2  \approx \frac{\omega_4}{\omega_4 \tau_3 +1} + \frac{1}{ \tau_1},
\end{equation}
which reproduces the condition for the decoherence free case.
  To obtain the position probability distribution, we integrate Eq.~(\ref{wigner})  over all $p$, which is a Gaussian integral.
Note that all terms proportional to $x^2$ cancel and we are left with an integral proportional to
\begin{equation}\label{in1}
 \langle x' | \rho_2(\tau_4) | x' \rangle  \propto \int d\Theta \frac{1}{\sqrt{a_1 - \frac{a_3^2}{4 a_2 }- 3 \I u_3 \tau_2 \Theta }} e^{\I \Theta x' + \I \frac{1}{4} \hbar^2 \tau_2 u_3 \Theta^3 - \left( \frac{1}{4 a_2} + \frac{\sigma_2^2}{2}  \right)\Theta^2},
\end{equation}
where we substituted $x$ with $x = \frac{ \omega_4 \tau_3 +1}{m \omega_4} \cosh(\omega_4 \tau_4) x' $ and where the index $2$ of $\rho_2(\tau_4)$ indicates that we considered decoherence only up to and including step 2 so far.
To proceed, consider the identity
\begin{equation}\label{dinga}
 \int d\Theta \frac{1}{\sqrt{\tilde{a}_1 - \I u_2 \Theta}} e^{\I x \Theta - \frac{1}{4 \tilde{a}_2} \Theta^2 + \I u_3 \Theta^3} \propto \left|\mathrm{Ai}\left( \frac{x}{\Delta x} \right) \ast e^{-\frac{x^2}{2 \hbar^2 \tilde{a}_1}}\right|^2,
\end{equation}
if $\tilde{a}_2 \tilde{a}_1 = 1/\hbar^2$, which resembles the Heisenberg limit for the harmonic oscillator ground-state. 
This identity can be proven by replacing $a_1$ with $\tilde{a}_1$, $a_2$ with $\tilde{a}_2$, setting $P(\tau_1) = 1$ and $\sigma_2 = 0$ in Eq.~(\ref{in1}) which would then assume an initial ground-state while ignoring all sources of decoherence. Thus, we can equate it to Eq.~(1) in the main text as both equations describe the exact same physics. Note that for an initial ground-state, $\tilde{a}_2 = (2 x_\mathrm{zp}^2)^{-1}$.
To turn Eq.~(\ref{in1}) in a suitable form to use the above identity, we add and subtract a term of the form $\frac{1}{4} \hbar^2 (a_1 - \frac{a_3^2}{4 a_2}) \Theta^2$ in the exponent.
After using the convolution theorem, re-substituting $x'$, and normalizing we obtain:
\begin{equation}\label{rho1}
\langle x | \rho_2(\tau_4) | x \rangle = \frac{1}{2 \pi \Delta x^2 \sigma_\mathrm{\Lambda2} \sigma_c} \left|\mathrm{Ai}\left( \frac{x}{\Delta x}\right) \ast e^{- \frac{x^2}{4 \sigma_c^2}} \right|^2 \ast e^{- \frac{x^2}{2\sigma_\mathrm{\Lambda2}^2}},
\end{equation}
with $\Delta x$ and $\sigma_c$ given by equations~(\ref{deltax}) and~(\ref{ss}) after replacing $\sigma_x(\tau_1)$ with $\sigma_x(\tau_1) = x_\mathrm{zp}^2(2 \bar{n} + 1) (\omega_0^2 \tau_1^2 + 1)$. The blurring variance excluding decoherence effects after step 2 is now given by
\begin{equation}\label{sigmal2}
    \sigma_\mathrm{\Lambda2}^2 =  \left[\left(\sigma_2^2 + \sigma_\mathrm{01}^2\right) \left(\frac{\omega_4 \tau_3 +1}{m \omega_4} \right)^2 \right] \frac{e^{ 2\omega_4 \tau_4}}{4},  
\end{equation}
with the associated momentum space blurring variance due to an initial thermal state (step 0) subject to displacement noise during the first free-evolution (step 1)
\begin{equation}\label{sigma01}
\sigma_\mathrm{01}^2=\frac{\hbar^2}{4} \frac{1-P^2(\tau_1)}{P^2(\tau_1) \sigma_x^2(\tau_1)}  =\frac{\hbar^2} {\sigma_x^2(\tau_1)} (\bar{n} +\bar{n}^2) + \frac{2}{3}  \hbar^2 \tau_1 \Lambda_1.
\end{equation}
Next we calculate the contributions to the blurring distance due to displacement noise during step 3 and 4.
\subsubsection{Displacement noise during Step 3 (second free-evolution) and during Step 4 (inverted potential)}
To derive the associated blurring distance due to displacement noise during step 3, we remember Eq.~(\ref{wig1})
\begin{equation}\label{wigb}
W_3(x,p,\tau_3) = \int dk_x dk_p e^{\I k_x x + \I k_p p} \tilde{W}_\mathrm{3,\Lambda_3=0}(k_x,k_p,\tau_3) \tilde{G}(k_x,k_p,\tau_3),
\end{equation}
where $\tilde{W}_\mathrm{3,\Lambda_3=0}(k_x,k_p,\tau_3)$ is the double Fourier transform of $W_2(x - (p/m) \tau_3,p,0)$, which is the Wigner function after the second free evolution including displacement noise during the first free evolution and displacement noise during step 2 but without displacement noise during the second free-fall as described by Eq.~(\ref{wigner}). 
To see how decoherence during the second free evolution translates into a blurring distance after step 4, we
again replace $x$ by $x_0$ with
\begin{equation}
 x_0 =x \cosh{(\omega_4 \tau_4)} - \frac{p}{m \omega_4}\sinh{(\omega_4 \tau_4)},
\end{equation}
and then all
$p$ by $p_0$ with
\begin{equation}
 p_0 =  p \cosh{(\omega_4 \tau_4)} -  m \omega_4 x \sinh{(\omega_4 \tau_4)}.
\end{equation}
Then we integrate over all $p$ since we are interested in the position probability distribution, 
and after integrating over $k_p$ and changing variables $y = k_x/\cosh(\omega_4 \tau_4)$ we obtain
\begin{equation}
\langle x | \rho_3(\tau_4) | x \rangle  \propto \int dy  e^{\I x y} \tilde{W}_\mathrm{\Lambda_3=0}( y \cosh(\omega_4 \tau_4), \frac{y}{m \omega_4} \sinh(\omega_4 \tau_4)  ) \tilde{G}(y \cosh(\omega_4 \tau_4), \frac{y}{m \omega_4} \sinh(\omega_4 \tau_4) )
\end{equation}
which after using the convolution theorem can be written as
\begin{equation}
\langle x | \rho_3(\tau_4) | x \rangle  \propto
\langle x | \rho_2(\tau_4) | x \rangle 
\ast e^{- \frac{x^2}{2 \sigma_3^2}},
\end{equation}
with 
\begin{equation}
 \bar{\sigma}_3^2 \approx \frac{2 \hbar^2 \Lambda_3 \tau_3^3}{3 m^2} \cosh^2(\omega_4 \tau_4),
\end{equation}
where we neglected the terms proportional to $\tau_3^2/\omega_4$ and $\tau_3/\omega_4^2$ as we assume $\omega_4 \tau_3 \gg 1$.
Thus, in order to account for displacement noise during the second free evolution we only have to add $\sigma_2^2$ to Eq.~(\ref{sigmal2}) by exploiting the associative property of convolutions.
\\
Displacement noise during step 4 is taken into account by another convolution with a Gaussian with associated blurring variance \cite{romero2017coherent}.
\begin{equation}
 \bar{\sigma}_4^2 = \frac{\hbar^2 \Lambda_4}{ 4 m^2 \omega_4^3} e^{2 \omega_4 \tau_4},    
\end{equation}
which is valid for $\omega_4 \tau_4 \gg 1$ and can be absorbed into the total blurring variance as well, which can be written as
\begin{equation}\label{sigmal}
    \sigma_\Lambda^2 =  \left[\left(\sigma_2^2 + \sigma_\mathrm{01}^2\right) \left(\frac{\omega_4 \tau_3 +1}{m \omega_4} \right)^2   + \sigma_3^2  + \sigma_4^2 \right] \frac{e^{ 2\omega_4 \tau_4}}{4} + \sigma_5^2, 
\end{equation}
with
\begin{equation}\label{sigma3}
    \sigma_3^2 = \frac{2\hbar^2 \Lambda_3 \tau_3^3}{3 m^2},
\end{equation}
and
\begin{equation}\label{sigma4}
    \sigma_4^2 = \frac{ \hbar^2 \Lambda_4}{ m^2 \omega_4^3}.
\end{equation}
 We also added $\sigma_5^2$ as the blurring variance due to the position detection itself which depends on the resolution of the implemented detection scheme. Note that  $\sigma_5^2$ can be neglected for sufficiently large $\omega_4 \tau_4$.
The final position probability distribution $P_D(x)$ (Eq.~(4) in the main text) is obtained by replacing $\sigma_\mathrm{\Lambda 2}$ by $\sigma_\Lambda$ in Eq.~(\ref{rho1}).

While the first moments of this distribution are unaffected by noise, the second moments in position and momentum are increased by adding a term of $\sigma_\Lambda^2$ and $m^2 \omega_4^2 \sigma_\Lambda^2$, respectively. These terms can be obtained by first calculating the moments of $|\Psi(x,\tau_4)|^2$ in the momentum picture (i), then writing the second convolution in integral form, solving the resulting Gaussian integral first and exploiting the result from (i).
\section{Black-body radiation}
In most experimental realisations of the suggested protocol, the dominant decoherence mechanism during the free evolutions (step 1 and step 3) is the emission and absorption of black-body radiation, which is typically dominated by the internal temperature of the particle. The internal temperature increases due to the absorption of black-body radiation emitted by the walls of the vacuum chamber and absorbed laser photons during laser light exposure. However, the internal temperatures decreases during the free evolutions by emitting black-body radiation into the environment. 

For the rest of this section, we assume an optical implementation. We begin by modeling the internal temperature of the particle during multiple experimental runs of our protocol.
We assume that groundstate cooling is performed in the intensity maximum of an optical standing wave for a cooling time $\tau_0$ leading to internal heating. Furthermore, we assume internal cooling during the rest of the protocol as most of the time the particle evolves freely without laser light exposure and only experiences negligible weak and short laser pulses with regard to the internal temperature of the particle during step 2 and 4.

The absorbed power from the trapping and cooling laser photons of a particle trapped in the anti-node of a standing wave can be derived to be  \cite{chang2010cavity,Ulbricht_near}
\begin{equation}
    P_\mathrm{abs} = 2 \pi \epsilon_0 E_0^2 k c r^3 \mathrm{Im} \frac{\epsilon_r(\omega_L)-1}{\epsilon_r(\omega_L) +2},
\end{equation}
where $E_0$ is the electric field amplitude of the standing wave, $k = 2 \pi/\lambda$ is the wavevector, $c$ the speed of light and $\epsilon_r(\omega_L)$ is the complex dielectric permittivity evaluated at the laser frequency $\omega_L$. 
$P_\mathrm{abs}$ can be expressed in terms the initial mechanical trapping frequency of the particle \cite{chang2010cavity}
\begin{equation}\label{omega}
 \omega_0 = \sqrt{\frac{3 k^2 \epsilon_0 E_0^2}{2 \rho} \mathrm{Re} \frac{\epsilon_r(\omega_L)-1}{\epsilon_r(\omega_L) +2}},
\end{equation}
where here $\rho$ is the particle density,
resulting in 
\begin{equation}
    P_\mathrm{abs} = \frac{m \omega_0^2 c}{k} \beta,
\end{equation}
with $\beta \equiv  \mathrm{Im} \frac{\epsilon_r(\omega_L)-1}{\epsilon_r(\omega_L) +2}/ \mathrm{Re} \frac{\epsilon_r(\omega_L)-1}{\epsilon_r(\omega_L) +2}$. We assume laser photons with a wavelength of $\lambda = 1550$~nm and use $\epsilon_r(\omega_L) = n_L^2$, with the complex refractive index of silica  $n_L(1550 \mathrm{nm}) = 1.44 + \I \, 2.5 \times 10^{-9}$ \cite{Ulbricht_near}.

\begin{figure} 
     \centering
\includegraphics[width=0.9\textwidth]{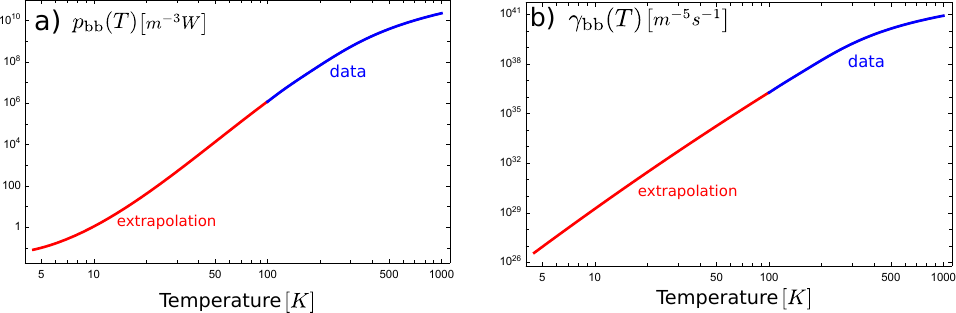}
\caption{\textbf{Black-body radiation: emitted/absorbed power and localization rate of silica nano-spheres)}\\
\textbf{a) emitted/absorbed power} $p_\mathrm{bb}(T)$ per unit volume as a function of temperature. Blue line is from integrating measurement data, red line is an extrapolation $p_\mathrm{bb}(T<100 \, K) \approx T^{-5.79} \exp\left(3.14 \ln^2(T)-0.265 \ln^3(T)  \right)$.  \\
\textbf{b) localization rate} $\gamma_\mathrm{bb}(T)$ per unit volume as a function of temperature. Blue line is from relating the emitted power in a) to the momentum recoil that the particle receives from the leaving photons. The red line is an extrapolation $\gamma_\mathrm{bb}(T<100 \, K) \approx 1.91 \times 10^{31} T^{8.38} \, \exp\left(-0.19 \ln^2(T) \right)$.
}  
 \label{Figsup1}
\end{figure}

With the internal particle energy $E_i = m c_m T_i$, where  $c_m = 700 \, \mathrm{J/(kg K)}$ is the specific heat capacity of silica, the differential equation for the internal temperature of the particle is given by
\begin{equation}\label{dgl}
 m c_m \frac{dT_i}{dt} = P_\mathrm{abs}(t) + V p_\mathrm{bb}(T_e) - V p_\mathrm{bb}(T_i),
\end{equation}
where $p_\mathrm{bb}(T)$ is the absorbed/emitted power per unit of volume $V$ of a silica particle. Its temperature dependency is numerically obtained by integrating the corresponding bulk measurement data \cite{Ulbricht_near} for $T>100$~K followed by calculating the power emitted by a thermal dipole and extrapolating it for $T < 100$~K as shown in Fig.~\ref{Figsup1}.
We assume that the particle has been trapped sufficiently long that it has reached a steady state internal Temperature $T_\mathrm{ss}$ before the first experimental run of the protocol, which is given by solving Eq.~(\ref{dgl}) for $T_i=T_\mathrm{ss}$, $P_\mathrm{abs}(t) = P_\mathrm{abs}$ and setting $dT_\mathrm{ss}/dt=0$. 
For a trapping frequency of $\omega_0 = 2 \pi \times 100$~kHz at $T_e = 300$~K, we obtain $T_\mathrm{ss} \approx 329 $~K. The reason why our calculated steady state temperature is much smaller as observed in optical tweezer traps is due to the much stronger curvature of the photon intensity mode profile (at the diffraction limit) in a standing wave setting. Thus, along a standing wave much lower laser powers are required to generate the same trapping frequencies as in optical tweezers. 

\begin{figure} 
     \centering
\includegraphics[width=0.9\textwidth]{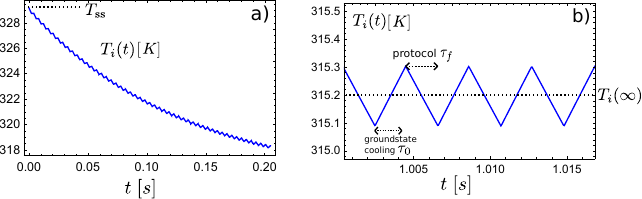}
\caption{\textbf{Internal Temperature $T_i(t)$: Case Study}\\
\textbf{a) Internal Temperature $T_i(t)$} as a function of time for $N = 50$ experimental runs for a standing wave trapping frequency of $\omega_0 = 2 \pi \times 100$~kHz at room temperature $T_e = 300$~K, cooling time of $\tau_0 = 2 $~ms and protocol time $\tau_f = 2.1$~ms\\
\textbf{b) The Internal Temperature $T_i(t)$} with same parameters as in a) reaches a dynamical steady state $T_i(\infty) \approx 315.2$~K after less than $N=244$ experimental runs.
}  
 \label{Figsup2}
\end{figure}

The above steady state temperature becomes the initial condition to simulate $N$ experimental runs with Eq.~(\ref{dgl}), where $P_\mathrm{abs}(t)$ is a piecewise function which starts with $0$ for time $\tau_f$, then $P_\mathrm{abs}(t)=P_\mathrm{abs}$ for time $\tau_0$, then $P_\mathrm{abs}(t)=0$ again for time $\tau_f$ and so on.
Note that the internal temperature does not depend on the particle Volume $V$ as all terms in Eq.~(\ref{dgl}) are proportional to $V$.
The solution to Eq.~(\ref{dgl}) with initial condition $T_\mathrm{ss} \approx 329 $~K  is shown in Fig.~\ref{Figsup2}a) for $N = 50$ experimental runs. As can be seen, in average, the internal temperature cools off until it reaches a dynamical steady state of $T_i(\infty) \approx 315.2$~K after less than $N = 244$ experimental runs as shown in Fig.~\ref{Figsup2}b), where heating during groundstate cooling and emission of black-body radiation during the free evolutions balance each other out. At this point the range of internal temperatures of the particle is just $\approx 0.2$~K. Hence, we use the time independent $T_i(\infty)$ for the localization rates describing the emission of black-body radiation during the free evolution and suggest that the measurement data obtained from the first 244 experimental runs should not contribute to the measurement of the position probability distribution.
Note that $T_i(\infty)$ can be calculated much faster with Eq.~(\ref{dgl}) by setting $dT_i/dt=0$ and multiplying $P_\mathrm{abs}$ with a factor of $\tau_0/(\tau_0 + \tau_f)$ and solving for $T_i=T_i(\infty)$.
This is a worst case approximation as the decohering effect of displacement noise scales with the position variance of the particle state, which is largest towards the end of the protocol, where the internal temperature drops below $T_i=T_i(\infty)$, as can be seen in Fig. \ref{Figsup2}b). Note that additional cooling time could in principle be added to the protocol if lower $T_i$ are required.
\subsubsection{Black-body localization rates}
For the black-body localization rates during the free-evolutions, the scattering of thermal photons can be neglected  \cite{romero2011quantum} and we are left with
photon emission and absorption:
\begin{equation}
 \Lambda_1 = \Lambda_3 = \Lambda_\mathrm{bb} = V \gamma_\mathrm{bb}(T_i(\infty)) + V \gamma_\mathrm{bb}(T_e).
\end{equation}
This expression can now be used to calculate the blurring distances $\sigma_\mathrm{01}$ (Eq.~\ref{sigma01}) during step 1 and $\sigma_3$ (Eq.~\ref{sigma3}) during step 3, which then get absorbed into the total blurring variance Eq.~(\ref{sigmal}).

Note that, the extrapolation of $\gamma_\mathrm{bb}(T)$ for $T < 100$~K is likely overestimating the actual localization rates since our extrapolation yields $\gamma_\mathrm{bb}(T \to 0) \gg 1$, as can be seen in Fig. \ref{Figsup1}b).

\section{Experimental Proposal: Optical Standing Wave}
Here we show how all required potentials can be generated by controlling the intensity and phase of an optical standing wave.
Then we proceed by deriving the associated photon recoil localization rates along an optical standing wave. In three dimensions, the normalized mode profile of an optical standing wave propagating in $x$-direction can be well approximated by $f(x,y,z) = \cos(k x -\phi) e^{-(y^2 + z^2)/w^2(x)}$, where $\phi$ is the controlled phase of the standing wave and $w(x)$ plays the role of an $x$-dependent beam waist. While the possibility of recoiled photons scattering into free-space requires a full 3D analysis, it is sufficient to focus on the direction of propagation ($x$) in order to derive the required potentials to implement the protocol.
\subsubsection{Optical Potentials along a standing wave}
The electric field amplitude of a standing wave propagating in $x$-direction can be written as
\begin{equation}
E(x) = E_0 \cos(kx- \phi),
\end{equation}
which generates an optical potential for dielectic nano- nano-particles of the form
\begin{equation}
V(x) = - \frac{m \omega^2}{2 k^2} \cos^2 \left( k \hat x - \phi \right),
\end{equation}
where $\omega$ is the mechanical frequency the particle experiences in the intensity maximum (see Eq.~\ref{omega}) and serves us in the following as a measure for laser intensity.

We obtain the required potentials for each step $i=0,1,3,4$ by controlling $\omega = \omega_i$  and the phase $\phi = \phi_i$, which stay constant during each step $i$.
As the particle should stay inside the trapping volume inside which it can be re-captured and groundstate cooled, we propose to approximately compensate arising linear optical potentials and the linear gravitational potential with an appropriate linear electric potential with opposite sign. The center of mass wave function of the particle evolves with the general Hamiltonian
\begin{equation}\label{ham}
    \hat H(\omega,\phi) = \frac{\hat p^2}{2m} + F(\omega,\phi) \hat x - \frac{m \omega^2}{2 k^2} \cos^2 \left( k \hat x - \phi \right),
\end{equation}
where
$
    F(\omega,\phi) \equiv \frac{m \omega^2}{2 k} \sin(2 \phi),
$ is implemented by a linear electric potential.
As $\phi$ and $\omega$ take different values for each step in our protocol, we shortly summarize the whole protocol:
\begin{itemize}
    \item Step 0: $H_0 = H(\omega_0,0)$, $\tau_0=0$, optically trapped and ground-state cooled nano-sphere at an anti-node of a standing wave.
    \item Step 1: $H_1 = H(0,0)$ free-evolution for time $\tau_1$.
    \item Step 2: $H_2 = H(\omega_p,\phi_2)$ for time $\tau_2$ which describes a short optical standing-wave pulse, which to leading orders contains a quadratic and a cubic potential.
    \item Step 3: $H_3 = H(0,0)$ second free-evolution for time $\tau_3$.
    \item Step 4: $H_4 = H(\omega_4,\pi/2)$ describes a rapid expansion in an inverted harmonic potential for time $\tau_4$ by shifting the standing wave such that the particle is at a node of the standing wave.
\end{itemize}
As step 0 describes the well known evolution in an harmonic potential, step 1 and 3 are free-evolution and step 4 generates the well-analysed inverted harmonic potential \cite{romero2017coherent, Pino2016}, we now focus on step 2.
\subsubsection{Short-lasting quadratic and cubic potential generated by optical standing wave pulse (step 2)}\label{short}
Expanding the potential in Eq.~(\ref{ham}) until third order around $x=0$ for an arbitrary phase $\phi_2$ yields:
\begin{equation}\label{cub}
    \hat H_2(\omega_p,\phi_2) \approx \frac{1}{2} m \omega_p^2 \cos(2 \phi_2) \omega_p^2  x^2 + \frac{1}{3} k m \omega_p^2
     \sin(2 \phi_2) x^3,
\end{equation}
where we dropped the constant part and assumed that the pulse is sufficiently short that we can neglect the kinetic term.
Before we discuss the short-pulse condition we want to focus on a couple of convenient properties of short pulses: 
\begin{itemize}
    \item As multiple subsequent short pulses commute with each other, the particle state does not depend on their order or if they are applied simultaneously. Thus, the quadratic and cubic pulse could also be applied separately in arbitrary order.
    \item Conveniently for experiments, only the pulse area matters (and not its shape): as $H(x,t) \approx V(x,t)$ commutes with itself at different times, the time evolution of a time dependent short-pulse Hamiltonian can be written as
\begin{equation}
 \Psi(x,t) = \exp\left(-\frac{\I}{\hbar} \int_0^t dt' V(x,t')\right) \Psi(x,0),
\end{equation}
Since $\int_0^t dt' V(x,t')$ is just proportional to the pulse area, we can in theory make the pulse arbitrary short and compensate the short duration with a larger peak intensity without affecting the final state. 
\end{itemize}
The last point enables us to always fulfil the conventional short-pulse requirement of $\langle \hat p^2 \rangle \tau_2 \ll 2 m \hbar$, especially in the case of relatively small pulse areas (as we will see are optimal with respect to decoherence). However, in reality the short-pulse approximation is much less restrictive if the short pulse is followed or proceeded by a free-evolution, as the first order correction is just an additional free-evolution for $\tau_2$ (which can be added to the followed or proceeded free-evolution time) and the next order correction is proportional to $\tau_2^2$.
To connect the Hamiltonian at step 2 (\ref{cub}) to the general form given in the main text, we identify:
\begin{equation}\label{omeg2}
    \omega_2^2 = \cos(2 \phi_2) \omega_p^2
\end{equation}
and
\begin{equation}
    \frac{1}{l} = \frac{1}{3} k \tan(2 \phi_2).
\end{equation}

\subsubsection{Photon recoil decoherence along an optical standing wave (step 0, 2 and 4)}
Here we derive the laser photon recoil induced localization rates
$\Lambda_r(\phi,\omega)$ along the direction of propagation of a standing wave. The dominant interaction process for a sub-wavelength particle is Rayleigh scattering. Localization happens due to the scattering of standing wave photons, which carry information about the center-of-mass position away. The Lindblad term can be derived to be \cite{nimmrichter2010master}
\begin{equation}\label{sca}
 L(\rho) = \gamma_\mathrm{sca} n_\mathrm{ph} \left[ \int d\Omega R(\vec{n}) f(\vec{r}) e^{- \I k \vec{n}\cdot \vec{r}} \rho e^{ \I k \vec{n}\cdot \vec{r}} f^*(\vec{r}) - \frac{1}{2} \{|u(\vec{r})|^2,\rho \}  \right],
\end{equation}
with the single photon Rayleigh scattering rate $\gamma_\mathrm{sca} = c k^4 \mathrm{Re}\left[\alpha(\omega_L)\right]^2/(6 \pi \epsilon_0^2 V_m)$, the particle polarizability $\alpha(\omega_L)$, the number of photons $n_\mathrm{ph} = E_0^2 \epsilon_0 V_m/(2 \hbar \omega_L )$, the mode volume $V_m$, the laser frequency $\omega_L$, the normalized mode profile $f(\vec{r}) = \cos(k x - \phi) e^{-(y^2 + z^2)/w^2(x)}$, and the emission pattern of a radiating dipole
$R(\vec{n}) = \frac{3}{8 \pi} (1- \sin^2(\Theta) \cos^2(\Phi))$  (here we assume the incident laser light is linearly polarized along the $y$-axis).
Assuming a motional state with a width $\sigma_\mathrm{x,y,z} \ll \lambda$, we approximate $w(x) \approx w(x_0) = \mathrm{const}$ and expand the terms
  $f(\vec{r}) e^{\pm \I k \vec{n}\cdot \vec{r}}$
until linear order around $x=0$, $y=0$, and $z=0$, which allows us to perform the integral with $d\Omega = \sin(\Theta) d\Phi d\Theta$:
\begin{equation}
 \int d\Omega R(\vec{n}) f(\vec{r}) e^{- \I k \vec{n}\cdot \vec{r}} \rho e^{ \I k \vec{n}\cdot \vec{r}} f^*(\vec{r}) \approx \beta_x(\phi) x\rho x + \beta_y(\phi) y \rho y + \beta_z(\phi) z \rho z + O\rho,
\end{equation}
where the term $O \rho$ is cancelled by a part of the last term of Eq.~(\ref{sca}), which maintains the Lindblad form of the master equation. Note that all cross-terms of the form $x \rho y$ vanish after integration.
The pre-factors are given by
\begin{align}
 & \beta_x(\phi) = \frac{1}{10} k^2 ( 7-3 \cos(2 \phi)  ) \\&
 \beta_y(\phi) = \frac{1}{5} k^2 \cos^2( \phi) \\ &
   \beta_z(\phi) = \frac{2}{5} k^2 \cos^2( \phi).
\end{align}
Note that $\beta_x + \beta_y + \beta_z = k^2$, implying that the total energy increase due to photon recoil does not dependent on $\phi$. However, the distribution of contributions to the localization rate for the different coordinates depends on the particle position along the standing wave. 
If the particle is at an anti-node ($\phi = 0$), we reproduce
the well known distribution of localization rates of $1/5$ (along polarization axis), $2/5$ (orthogonal to direction of propagation and polarization axis), $2/5$ (direction of propagation) for a far-detuned atom trapped in the intensity maximum of a standing wave \cite{wineland1979laser}.
Counter-intuitively, for a particle in a node $\beta_x(\pi/2) = k^2$ is maximal, while $\beta_y(\pi/2) = \beta_z(\pi/2) = 0$, consistent with the fact that in a node the quadratic potential orthogonal to the direction of propagation vanishes.
The photon recoil localization rates along a standing wave can be expressed as
\begin{equation}\label{lambdar}
    \Lambda_r(\phi,\omega) = \frac{\pi^2 \omega^2 \rho V^2}{5 \hbar \lambda^3} \mathrm{Re}\left[\frac{\epsilon_r(\omega_L)-1}{\epsilon_r(\omega_L)+2} \right] (7-3 \cos(2 \phi)),
\end{equation}
where $\omega$ is the trapping frequency the particle would experience at an anti-node ($\phi=0$) and which reproduces rates previously calculated for a particle actually trapped at an anti-node \cite{chang2010cavity}.
Thus, for step 2, $\Lambda_2 = \Lambda_r(\phi_2,\omega_p)$,
where $\phi_2$ and $\omega_p$ remain to be optimized and for step 4, $\Lambda_4 = \Lambda_r(\pi/2,\omega_4) = \frac{2 \pi^2 \omega_4^2 \rho V^2}{\hbar \lambda^3} \mathrm{Re}\left[\frac{\epsilon_r(\omega_L)-1}{\epsilon_r(\omega_L)+2} \right]$. These expressions can now be used to specify the previously calculated blurring distances $\sigma_2$ (Eq.~(\ref{sigma2})) and $\sigma_4$ (Eq.~(\ref{sigma4})), which are then absorbed into the total blurring distance $\sigma_\Lambda$ (Eq.~(\ref{sigmal})).

\section{Gas Collisions}
Since during our protocol the thermal de-Broglie wavelength $\lambda_\mathrm{th} = \sqrt{2 \pi \hbar^2/(m_g k_b T_e)}$ of the background gas molecules with mass $m_g$ and temperature $T_e$ is much smaller than the position width of the particle state $\sigma_x$, a single gas collision localizes the state. Thus, to have a negligible effect of gas collisions on the measured interference pattern, most experimental runs should be completely free of gas collisions.
Here, we first derive the probability $\mathcal{P}_0(T_e, r,P,\tau_f)$ of having no gas collisions during an experimental run lasting $\tau_f$, which depends on the particle size $r$, the temperature $T_e$ and the pressure $P$ inside the vacuum chamber. In the second part we provide the required pressures $P_\mathrm{0.9}$ such that 90\% of experimental runs are free of gas collisions which could resolve the position $x$ or transfer momentum along the $x$-direction.

Assuming a Maxwell-Boltzmann distribution for the velocities of the gas particles, we arrive at an average collision rate of
\begin{equation}
    \gamma_g = \frac{8 \pi P r^2}{m_g \langle v_g \rangle},
\end{equation}
where $P$ is the pressure in the vacuum chamber and $\langle v_g \rangle = \sqrt{8 k_b T_e/(\pi m_g)}$ is the mean velocity of the Maxwell-Boltzmann distribution of the gas molecules. Assuming gas-particle collisions are statistically independent, rare and $\gamma_g$ does not depend on time, it can be shown that the collisions follow Poissonian statistics and the probability of $n$ gas collisions during the time $\tau_f$ is given by
\begin{equation}
    \mathcal{P}_n(T_e,r,P,\tau_f) = \frac{(\gamma_g \tau_f)^n}{n!} e^{- \gamma_g \tau_f}.
\end{equation}
As we are only interested in the $x$-direction, the probability of no gas collisions with a momentum-component along this single spacial degree of freedom is
\begin{equation}
    \mathcal{P}_x(T_e,r,P,\tau_f) = e^{- \gamma_g \tau_f/3},
\end{equation}
where we assume that the scattering rate along the $x$-direction is $\gamma_g/3$.
\begin{figure} 
     \centering
\includegraphics[width=0.9\textwidth]{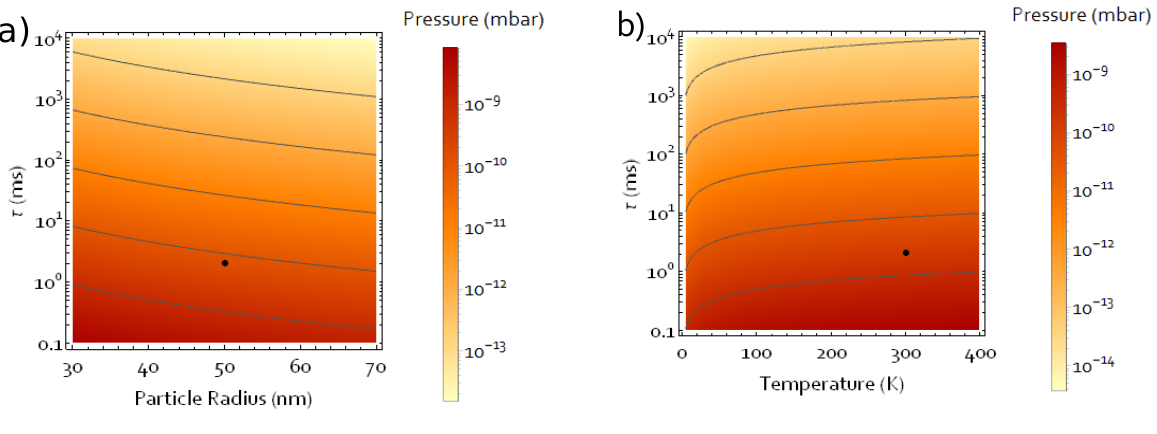}
\caption{\textbf{Required Pressure $P_\mathrm{0.9}$} to ensure that 90\% of experimental runs are gas collision free along the $x$-direction by enforcing $ \mathcal{P}_x(Te,r,P_\mathrm{0.9},\tau_f) = 0.9$. \\
\textbf{a)} Required pressure $P_\mathrm{0.9}$ as function of particle radius $r$ and total protocol time $\tau_f$ for a fixed environmental temperature $T_e = 300$~K.\\
\textbf{b)} Required pressure $P_\mathrm{0.9}$ as function of environmental temperature $T_e$ and total protocol time $\tau_f$ for a fixed  particle radius $r=50$~nm. \\
In both plots we assume that the gas composition is dominated by $H_2$ molecules with $m_g = 2$~u. The black dot represents our case study, which requires a pressure of $P_\mathrm{0.9} \approx 1.4 \times 10^{-10}$~mbar.
}  
 \label{Figsup3}
\end{figure}
Since we operate in ultra high vacuum, we assume a gas composition dominated by $H_2$ molecules with $m_g = 2$~a.m.u. Furthermore, we fix $ \mathcal{P}_x(T_e,r,P_\mathrm{0.9},\tau_f) = 0.9$, ensuring that 90\% of experimental runs are gas collision free along $x$. Fig.~\ref{Figsup3}a) shows the required pressures $P_\mathrm{0.9}$ in order to ensure $ \mathcal{P}_x(T_e=300K,r,P_\mathrm{0.9},\tau_f) = 0.9$ as a function of particle radius $r$ and protocol time $\tau_f$ in a room temperature environment. Fig.~\ref{Figsup3}a) shows the required pressures $P_\mathrm{0.9}$ in order to ensure $ \mathcal{P}_x(T_e,r=50\mathrm{nm},P_\mathrm{0.9},\tau_f) = 0.9$ as a function of environmental temperature $T_e$ by fixing the particle radius to $r = 50$~nm.
The black dot in both plots represents our case study where we assume a silica particle with $r = 50$~nm, which requires a pressure of $P_\mathrm{0.9} \approx 1.4 \times 10^{-10}$~mbar at room temperature in order experience no gas collisions in 90\% of experimental runs with $\tau_f \approx 2.1$~ms.

\section{Coherence: Quantification and Certification}
As metric for the degree of coherence between two arbitrary positions $x_1$ and $x_2$ of a given particle density matrix $\rho_\mathrm{ij} = \langle x_i | \rho | x_j \rangle$ we use  the first order correlation function $g_1(x_1,x_2,t)$ \cite{glauber1963quantum} by replacing electric fields with density matrix elements at a given moment in time:
\begin{equation}
    g_1(x_1,x_2,t) = \frac{\rho_\mathrm{12}(t)}{\sqrt{\rho_\mathrm{11}(t) \rho_\mathrm{22}(t)}},
\end{equation}
where in the case of $\rho_\mathrm{11} = \rho_\mathrm{22}$ (equal amplitudes), $|g_1(x_1,x_2)|$ is the visibility of an interference pattern created with a perfect double-slit experiment (infinitely narrow slits, located at $x_1$ and $x_2$, no added decoherence). 
For a Gaussian state, 
\begin{equation}
    |g_1(x_1,x_2,t)| = \exp \left(- \frac{(x_1-x_2)^2}{2 x_c^2(t)} \right),
\end{equation}
is just another Gaussian. With a given purity $P(t) = \mathrm{Tr(\rho^2(t))}$ at time $t$ and position width $\sigma_x(t)$, we call the length-scale 
\begin{equation}
    x_c(t) = \frac{2 P(t) \sigma_x(t)}{\sqrt{1 - P^2(t)}},
\end{equation}
over which the first order correlation function decays (its standard deviation) "coherence length" (in analogy with optical coherence), which quantifies the spatial distance over which positions are predominantly in a coherent superposition (have a fixed phase relationship) and could in principle destructively interfere with each other. This definition implies that for positions $|x_2-x_1|> x_c$ (which are further apart than $x_c$), the first order correlation function $g_1(x_1,x_2) < \exp(-0.5) \approx 0.6$.

Interestingly, the contribution $\sigma_\mathrm{01}^2$ to the total blurring variance $\sigma_\Lambda^2$ caused by an impure state $P(\tau_1)<1$ after the first free evolution (step 1) is just inversely proportional to $x_c^2(\tau_1)$ times the square of the classical momentum to position mapping function as can be seen in Eq.~(\ref{sigma01}), where $\sigma_\mathrm{01}^2 = \hbar^2/x_c^2(\tau_1)$.
One can extract $\sigma_\Lambda$ by de-convolving the final interference pattern until the result has maximum visibility and a lower bound on $x_c(\tau_1)$ can be found by
pessimistically assuming that all blurring was caused by a finite $x_c(\tau_1)$ at step 1.
Thus, $x_c(\tau_1) > \frac{\hbar}{\sigma_\Lambda} x_\mathrm{map}(t)$,
where $x_\mathrm{map}(t)$ is the classical momentum to position space mapping function.
However, with a cubic phase-grating protocol it is possible to certify a lower bound on $x_c(\tau_1)$ without knowing $x_\mathrm{map}(t)$ by also extracting $\sigma_c$ from the final interference pattern and measuring the position variance $\sigma_x^2(\tau_1)$ after step 1. 
Calculating  $\sigma_c/\sigma_\Lambda$, the classical mapping function cancels and we obtain:
\begin{equation}\label{lowerbound}
    x_c(\tau_1) > 2 \sigma_x(\tau_1) \frac{\sigma_c}{\sigma_\Lambda}.
\end{equation}
Note that for this result we don't need to know any additional experimental parameters.

\section{5$\sigma$ confirmation of the particle wave-nature: A Toy-Model} 
 Here, we address the confidence we can have in the observation of the wave-nature of the particle based on our measurement data. The signature of detection is the observation of an interference pattern, which implies the existence of a local minima surrounded by two interference peaks. We focus on the difference in detection events between the minimum surrounded by the first and the second largest interference maximum (see Fig~2 in the main text). For this toy model, we only address the difference in detection events between the \textit{second} largest peak and the minimum.
We will now estimate the experimental runs (ensemble size) required to confirm their difference in detection events with $m \times \sigma$ confidence. A more sophisticated evaluation is likely to reduce the number of required experimental runs further.
 
\begin{figure} 
     \centering
\includegraphics[width=0.9\textwidth]{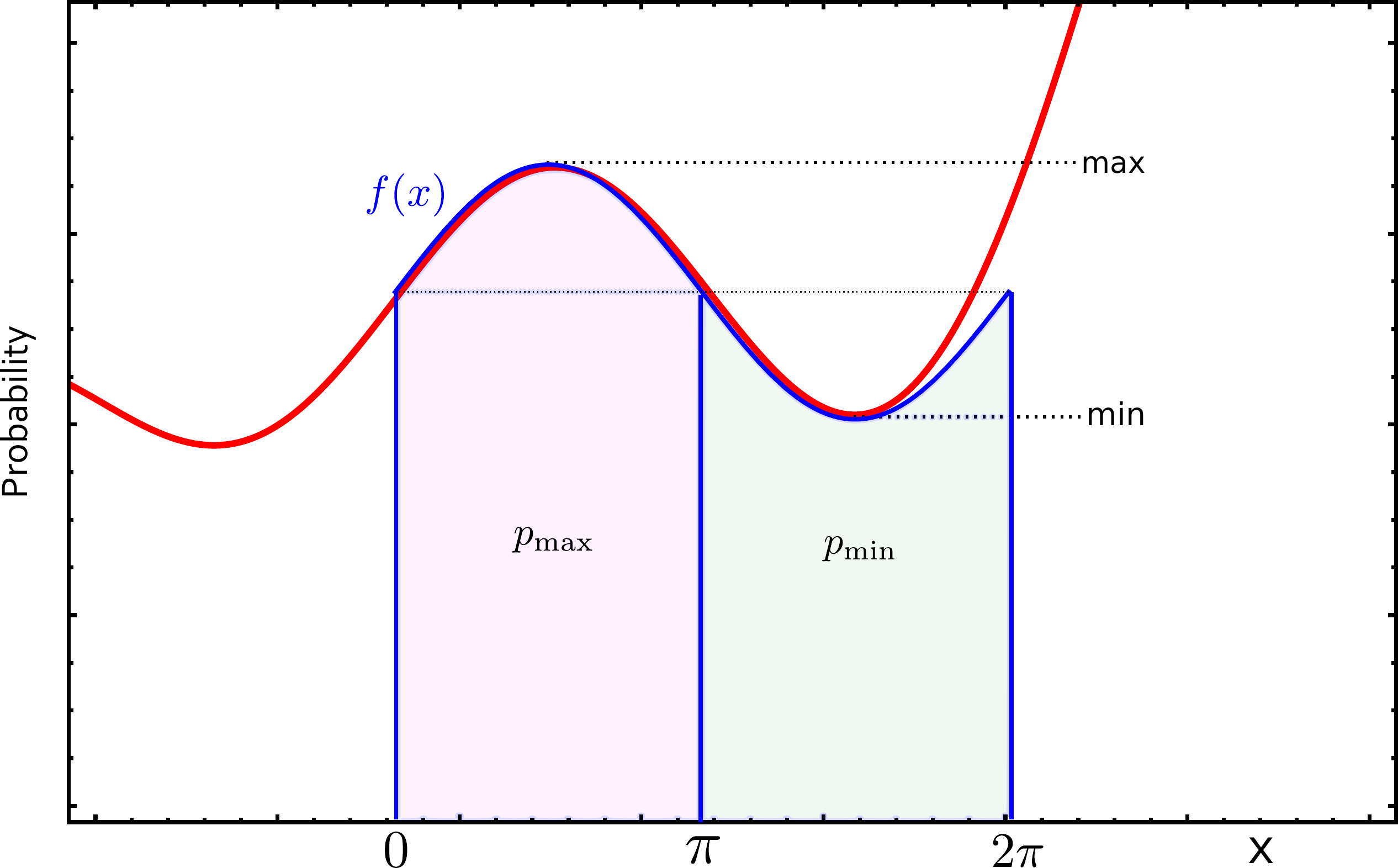}
\caption{\textbf{Experimentally distinguishing a minimum from a maximum.} 
The relevant area of the final interference pattern $P_D(x)$ (red) shown in Fig.~2 (main text) can be roughly fitted to a sinusoidal function $f(x)$ (blue, Eq.~(\ref{fx})), such that a single oscillation captures the relevant region $0 \le x \le 2\pi$. We divide the relevant area $p_r \approx  \pi (\mathrm{max}+ \mathrm{min}) = p_\mathrm{max} + p_\mathrm{min} = \int_0^{\mathrm{2 \pi}} dx f(x)$ into two slots, where $p_\mathrm{max}$ and $p_\mathrm{min}$ are the probabilities of finding the particle in the area which contributes to the maximum and minimum, respectively.
}  
 \label{Figsup4}
\end{figure}

 
 To simplify the estimation, we roughly fit
 the relevant region (second largest peak and first minimum) of the final interference pattern $P_D(x)$ (Fig.~2 and Eq.~(4) in main text) to a full period of a sinusoidal function
 \begin{equation}\label{fx}
     f(x) = \frac{1}{2}(\mathrm{max}-\mathrm{min}) \sin(x) + \frac{1}{2}( \mathrm{max} + \mathrm{min}),
 \end{equation}
 which is illustrated in Fig.~\ref{Figsup4}. Here, we absorb probability normalization into the values of $\mathrm{max}$ and $\mathrm{min}$ and only care about the end result and its scalings.
The probability $p_r = p_\mathrm{max} + p_\mathrm{min}$ of finding the particle at the relevant positions (area below $f(x)$) is divided into two detection slots $p_\mathrm{max}$ and $p_\mathrm{min}$, which are the probabilities of detecting a particle contributing to the maximum or minimum, respectively, per experimental run.
 For $N$ experimental runs, the number of detection events in each slot are then given by $n_\mathrm{max} \approx N \int_0^\pi dx f(x) $ and $n_\mathrm{min} \approx N \int_\pi^{2\pi} dx f(x) $.
 To confirm the existence of a minimum with $m \times \sigma$ confidence, the difference in detection events inside each slot should be $m$ times larger than the combined standard deviation of each slot, which is mathematically expressed with the condition
 \begin{equation}
     n_\mathrm{max}-n_\mathrm{min} \gtrapprox m \sqrt{\sigma^2_\mathrm{max}+\sigma^2_\mathrm{min}},
 \end{equation}
 with statistical uncertainties $\sigma^2_\mathrm{max} =  n_\mathrm{max}$
 and $\sigma^2_\mathrm{min} =  n_\mathrm{min}$.
 With the definition of the visibility $v = (\mathrm{max}-\mathrm{min})/(\mathrm{max}+\mathrm{min})$, we arrive at the condition
 \begin{equation}\label{Nexp}
     N \gtrapprox \frac{\pi^2}{4} \frac{m^2}{v^2 p_r},
 \end{equation}
 in order to obtain a $m \times \sigma$ confirmation of the particle wave-nature. To conveniently apply this formula to Eq.~(4) in the main text we use
 \begin{equation}
     \left|\mathrm{Ai}\left(\frac{x}{\Delta x} \right) \ast \exp \left( - \frac{x^2}{4 \sigma_c^2} \right)\right|^2 \propto \left|\mathrm{Ai}\left(\frac{x}{\Delta x} + \frac{\sigma_c^4}{\Delta x^4} \right)\right|^2, 
 \end{equation}
 note that the subsequent convolution with $\exp \left(- x^2/(2 \sigma_\Lambda^2) \right)$ has no effect on the location of the minima and maxima and then numerically calculate the corresponding maxima and minima of $\mathrm{Ai}^2(x)$. 
 Then, 
 \begin{equation}
     \mathrm{max} \approx P_D(-3.248 \Delta x - \sigma_c^4/\Delta x^3),
 \end{equation}
 and
 \begin{equation}
     \mathrm{min} \approx P_D(-2.338 \Delta x - \sigma_c^4/\Delta x^3).
 \end{equation}
 Furthermore, $p_r$ is roughly the area below the second largest peak:
 \begin{equation}
     p_r \approx \int_{x_\mathrm{min2}}^{x_\mathrm{min1}} dx P_D(x),
 \end{equation}
 with
 \begin{equation}
     x_\mathrm{min2} \approx -4.088 \Delta x - \sigma_c^4/\Delta x^3, 
 \end{equation}
 and
  \begin{equation}
     x_\mathrm{min1} \approx -2.338 \Delta x - \sigma_c^4/\Delta x^3 , 
 \end{equation}
being the two minima surrounding the second largest peak.
  For Fig.~3 and Fig.~4 in the main text, we fix $N \approx 1.2 \cdot 10^4$ and $m = 5$ for each data point allowing different combinations of $v$ and $p_r$ which can be optimized for any specific goal.
 
\section{Optimization of experimental parameters for Fig.~3 and Fig.~4}
Here we describe a fast way to find the optimal combination of experimental parameters $\omega_i$, $t_i$ and $\phi_2$ which maximize the lower bound on certified coherence length $x_c(\tau_1)$ (Fig.~3) or coherent peak distance $\propto \Delta x(\tau_3) $ (Fig.~4).
In both figures, we set a "quality-standard" for the final interference pattern, which we enforce for all data points. We achieve this by roughly fixing the number of experimental runs required to have a 5$\sigma$ confirmation of matter-wave interference, which comes down to fixing the quality-parameter 
\begin{equation}\label{quality}
   q \equiv v^2 p_r \approx (5 \pi/2)^2 N_\mathrm{5\sigma}^{-1} 
\end{equation}
for each data point in the figures.

\subsection{Enforcing a quality standard}
  Remarkably, writing $x$ in Eq.~(4) (main text) in units of $\Delta x$ leads to an equation, which only depends on two parameters: $p_c \equiv \sigma_c/\Delta x$ and $p_\Lambda \equiv \sigma_\Lambda/\Delta x$. We call these two parameters the "shape-parameters" since their values completely determine the shape and thus also the visibility $v$, relevant area $p_r$ and therefore the quality $q = v^2 p_r$ of the final interference pattern. However, different values of $p_c$ and $p_\Lambda$ can lead to the same quality $q$ since a lower visibility $v$ can be compensated by a larger relevant area $p_r$ (or vice versa). Their optimal combination depends on the metric we wish to maximize. 
    \begin{figure} 
     \centering
\includegraphics[width=0.9\textwidth]{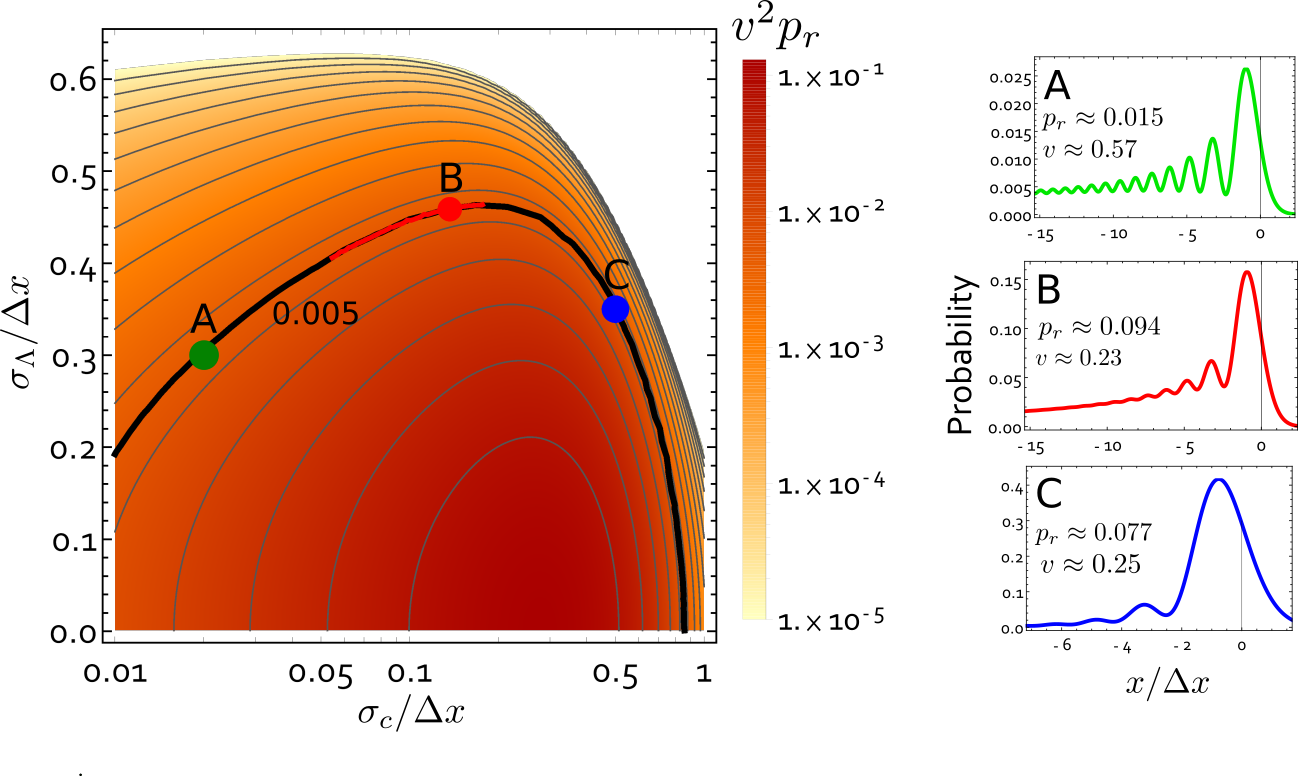}
\caption{\textbf{Shape-Space: The space of possible shapes of interference patterns.}\\
Left: Shape-quality $q = v^2 p_r$ as a function of shape-parameters  $p_c \equiv \sigma_c/\Delta x$ and $p_\Lambda \equiv \sigma_\Lambda/\Delta x$. In the white region $v = 0$. The black contour fixes $q=0.005$, which corresponds to $N_\mathrm{5\sigma} \approx 1.2 \cdot 10^4$. The points A, B and C are example shapes, which are shown on the right side with their corresponding values of $p_r$ and $v$. The region around B, which also represents our case study with $p_c \approx 0.14$, tolerates the largest amount of noise (quantified by $\sigma_\Lambda$). The red line on top of the black line shows all possible accessible shapes for our case study with a fixed total protocol time of $\tau_f \approx 2.1$~ms. To produce the other possible shapes, $\phi_2$ and $\tau_1$ need to be chosen accordingly, while the other parameters are the same as in the example parameter set (table~1) in the main text. The point B maximises the lower bound of $x_c(\tau_1)$ of all accessible shapes. All optimal combinations of parameters for Fig.~3 and 4 in the main text are roughly found between A and B as for both plots large $\sigma_c/\Delta x$ lead to lower values of $x_c(\tau_1)$ or $\Delta x$, respectively.
}  
 \label{Figsup5}
\end{figure}

We enforce the chosen quality-standard $q$ by restricting the search space to shape-parameters $p_c$ and $p_\Lambda$ which satisfy the contour-equation
\begin{equation}\label{cont1}
    q(p_c,p_\Lambda) = v^2 p_r \approx (5 \pi/2)^2 N_\mathrm{5\sigma}^{-1} = \mathrm{const},
\end{equation}
where we fix $N_\mathrm{5\sigma}$ as we desire. In Fig.~\ref{Figsup5} (left) we plot the shape-quality $q$ as a function of $p_c$ and $p_\Lambda$ which span up the shape-space, which is the space of all possible shapes. On the right, we show three examples of possible shapes for $q=0.005$ (black line) which corresponds to $N_\mathrm{5\sigma} \approx 1.2 \cdot 10^4$. 
Once we decide on a wished shape-quality, we numerically solve
the contour-equation (\ref{cont1}) for
\begin{equation}
    p_\Lambda^{(q)}\left[p_c\right], 
\end{equation}
 which tells us the required $p_\Lambda$ for a given $p_c$ to meet our quality-standard $q$. 
  \subsection{Enforcing the condition Eq.~(\ref{cond})}
 Enforcing the condition (\ref{cond}) allows us to eliminate $\omega_2^2 \tau_2$ in $p_c$ and $p_\Lambda$ by replacing it with $\tau_1^{-1} + \tau_3^{-1}$, where we assume $\omega_4 \tau_3 \gg 1$.
Note that in $p_\Lambda$, the product $\omega_2^2 \tau_2$ appears in
$\sigma_2$ (Eq.~(\ref{sigma2})) after replacing $\Lambda_2$ with Eq.~(\ref{lambdar}) and $\omega_p^2$ with Eq.~(\ref{omeg2}).

\subsection{Choosing values for $\omega_0$ and $\omega_4$}
As quality $q = v^2 p_r$ monotonically increases with larger values of $\omega_0$ and $\omega_4$, these values are constrained by experimental boundary conditions. Here we choose
$\omega_0 = 2 \pi \times 100$~kHz to a experimentally realistic initial trapping frequency compatible with previous ground-state cooling experiments. However, larger values of $\omega_0$ significantly decrease fringe blurring caused by a finite thermal occupation (Eq.~(\ref{sigma01})) allowing us to tolerate larger initial $\bar{n}$.

For $\omega_4 \tau_3 \gg 1$,  $\omega_4$ only appears in the blurring variance $\sigma_4^2 \propto \omega_4^{-1}$.  Here, we fix $\omega_4 = 2 \pi \times 10$~kHz as the blurring caused by the cubic pulse is roughly $\omega_4 \tau_3$ larger than the blurring caused by the inverted potential and further increasing $\omega_4$ does not significantly increase $q$. This choice implies $\tau_3 \gg 0.016$~ms to satisfy the assumption $\omega_4 \tau_3 \gg 1$.

\subsection{Enforcing a fixed fringe distance in Fig.~(3) (with inverted potential)}\label{sec:fringe}
In Fig.~(3) (main text), all data points have the same
distance between the first two interference minima $\approx 1.75 \Delta x = 5$~nm. 
Remarkably, the inverted potential effectively decouples the shape ($p_c$,$p_\Lambda$) from the size $\propto \Delta x$ of the final interference pattern, since $p_c$ and $p_\Lambda$ do not depend on $\tau_4$, while $\Delta x \propto \exp(\omega_4 \tau_4)$.
Hence, as $\tau_4$ has no effect on shape and quality, we fix the distance between the first two interference minima by simply solving the equation $1.75 \Delta x = 5$~nm for $\tau_4$ after finding the optimal combination of the remaining free parameters, which are $\tau_1$, $\tau_3$ and $\phi_2$. 

\subsection{Maximizing the lower bound on coherence length $x_c(\tau_1)$ for Fig.~3 in the main text}
Now, $p_c$ is a function of $\tau_1$, $\tau_3$, $\phi_2$ and $r$, while $p_\Lambda$ is a function of $\tau_1$, $\tau_3$, $\phi_2$, $r$, $T_e$ and $T_i$. The total protocol time is dominated by the two free evolutions $\tau_f \approx \tau_1 + \tau_3$ (which is on the order of milliseconds) since $\tau_4 \sim 0.1$~ms and $\tau_2$ can be chosen arbitrarily short as only pulse area counts (here we choose $\tau_2 = 10^{-2}$~ms). Thus, for Fig.~3, we first calculate $T_i = T_i(\infty)$ for a given $\tau_f$ and $T_e$ with Eq.~(\ref{dgl}) by choosing a groundstate cooling time of $\tau_0 = 2$~ms. Then, we fix a particle radius $r$ and numerically find the combinations of $\tau_1$ (note that $\tau_2 = \tau_f - \tau_1)$) and $\phi_2$ which solve the contour-equation  
 \begin{equation}\label{cont2}
    p_\Lambda^{(q)}\left[p_c(\tau_1,\phi_2)\right] = p_\Lambda(\tau_1,\phi_2],  
\end{equation}
for a given $r$, $\tau_f$ and $T_e$, which guaranties our chosen quality standard. The range of possible values of $p_c$ and $p_\Lambda$ with these found combinations of $\tau_f$ and $\phi_2$ for $\tau_f \approx 2.1$~ms, are drawn into Fig.~\ref{Figsup5} (red line around point B), where the remaining parameters are as in the example parameter set in the main text. 
Next, we plug all found combinations of $\tau_1$ and $\phi_2$ (if there are any) into the equation for the lower bound of $x_c(\tau_1)$ (Eq.~(\ref{lowerbound})) and pick the combination which yields the largest value (Point B in Fig.~\ref{Figsup5}). All optimal shapes for Fig.~3 in the main text are roughly found between A and B as large $\sigma_c/\Delta x$ and simultaneously small $\sigma_\Lambda/\Delta x$ imply a small $\tau_1$. A small $\tau_1$ requires large $\omega_2^2 \tau_2$ to meet the condition~(\ref{cond}), which increases $\sigma_\Lambda/\Delta x$ leading to lower values of $x_c(\tau_1)$.

With the combination of $\tau_1$ and $\phi_2$, which maximizes the lower bound of $x_c(\tau_1)$, we calculate the required $\tau_4$ in order to have $5$~nm fringe spacing as explained in section \ref{sec:fringe}. On the y-axis in Fig.~3 we show $\tau_f = \tau_f + \tau_4 + \tau_2$ and plot the largest found value of $x_c(\tau_1)$.
In Fig.~3a) we fix $r= 50$~nm and scan $T_e$ for the x-axis, while in Fig.~3b) we fix $T_e = 300$~K and scan $r$ for the x-axis. 
\subsection{Maximizing coherent peak distance for Fig.~4 in the main text}
First we calculate $g_1(x_\mathrm{max1},x_\mathrm{max2},\tau_3)$ (defined in the next section) for all combinations of $p_c$ and $p_\Lambda$ which satisfy the contour-equation (\ref{cont1}) for $N_\mathrm{5\sigma} \approx 1.2 \cdot 10^4$. Since for all combinations $g_1(x_\mathrm{max1},x_\mathrm{max2},\tau_3) > 0.95$, we conclude that our applied quality-standard implies coherence of $>0.95$ between the two largest peaks as long as additional decoherence during the second free evolution can be neglected.
Then we do the exact same thing as in the previous section but instead of picking the largest $x_c(\tau_1)$, we pick the largest peak distance $\approx 2.23 \Delta x(\tau_3)$ from the found combinations of $\tau_1$ and $\phi_2$ which satisfy Eq.~(\ref{cont2}). Here we replace
$\Delta x$ with $\Delta x(\tau_3)$, $\sigma_c$ with $\sigma_c(\tau_3)$ and $\sigma_\Lambda$ with $\sigma_{\Lambda_3}$, which are defined in the following section.  

\section{Coherent Splitting without inverted potential}
\begin{figure} 
     \centering
\includegraphics[width=0.6\textwidth]{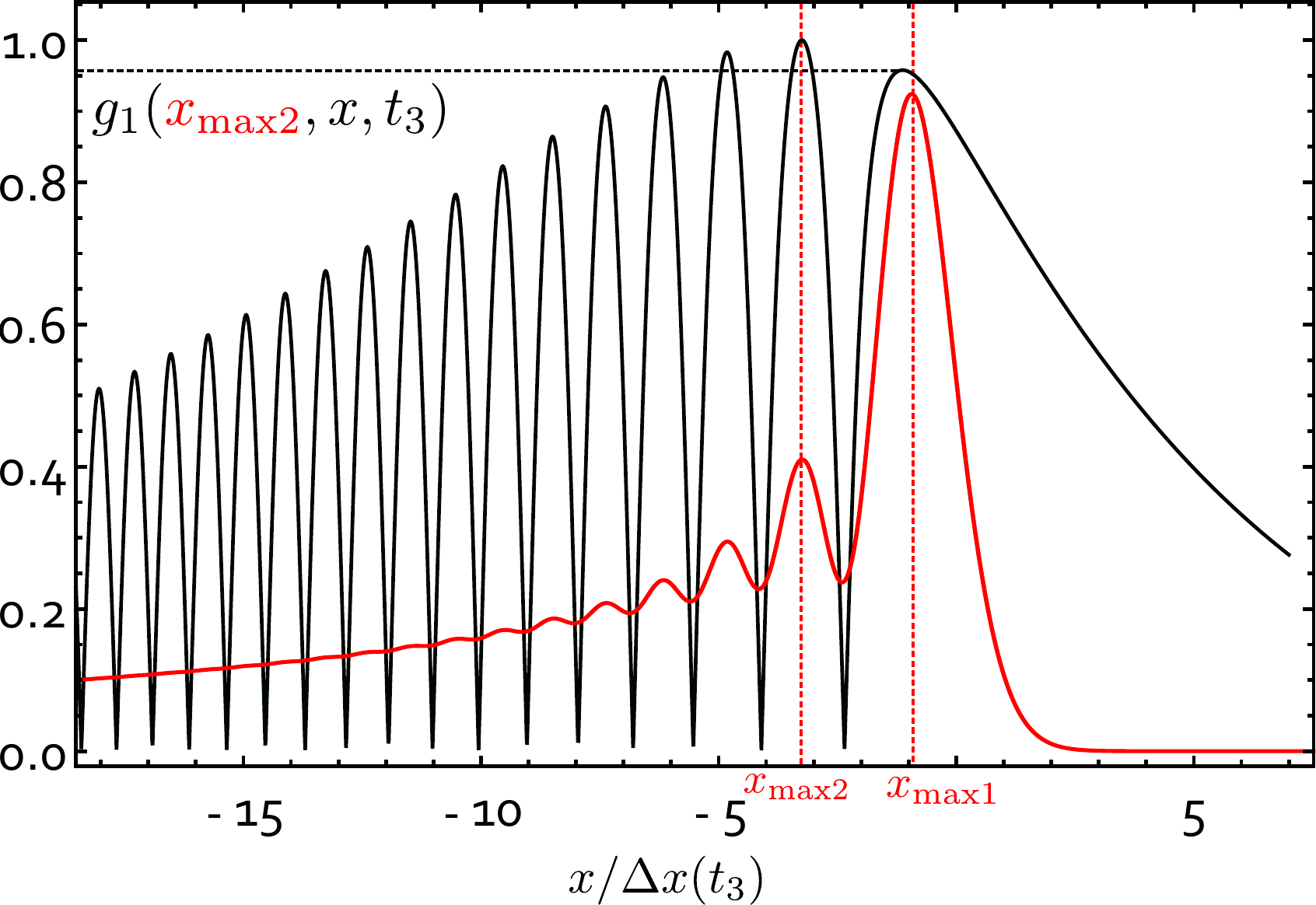}
\caption{\textbf{Coherence between interference peaks.}
Shown is $g_1(x,x_\mathrm{max2},\tau_3)$ (black) as a function of $x$, which quantifies coherence between the second largest peak $x_\mathrm{max2}$ and an arbitrary position $x$. For $x = x_\mathrm{max1}$ (the largest peak), we observe
$g_1(x_\mathrm{max1},x_\mathrm{max2},\tau_3) \approx 0.96$. The interference pattern (red) corresponds to Fig.~4 in the main text, with peak distance $x_\mathrm{max2}-x_\mathrm{max1} \approx 2.23 \Delta x(\tau_3) = r = 50$~nm which requires a total protocol time of $\tau_f \approx 350$~ms with $\tau_1 \approx 0.92$~ms, $\tau_3\approx 349$~ms, $\phi_2 \approx 0.9 \pi/4$, $\omega_2 \approx 2 \pi \times 1.66$~kHz. We assume $\bar{n}=0.5$. The probability of finding the particle within the two largest peaks is $\approx 0.31$.
}  
 \label{Figsup6}
\end{figure}

Here we are interested in the coherence between individual peaks of the final position probability distribution. We skip step 4 (inverted potential), as we estimate the requirements on localization rates in order to maintain significant coherence as hard or impossible to achieve with existing technology. However, clean electric potentials might change that in the future. We consider an optical standing wave implementation in a cryogenic environment and thus only take photon recoil decoherence and an initial thermal state into account.The requirements on pressure and temperature can then be extracted from the total protocol time by calculating the corresponding gas scattering rates and black-body localization rates. 
To quantify the coherence between different positions, we now proceed by calculating the first order correlation function 
\begin{equation}
    g_1(x_1,x_2,\tau_3) = \frac{|\langle x_1|\rho(\tau_3) | x_2 \rangle|}{\sqrt{\langle x_1|\rho(\tau_3) | x_1 \rangle \langle x_2|\rho(\tau_3) | x_2 \rangle}},
\end{equation}
after $\tau_3$ (after step 3).
It is convenient to substitute $x_1 = x + x'/2 $ and $x_2 = x - x'/2$, as arbitrary density matrix elements of $\rho(\tau_3)$ can be calculated with
\begin{equation}
\langle  x + \frac{x'}{2}| \rho(\tau_3) | x - \frac{x'}{2} \rangle = \frac{1}{h} \int dp e^{\I x' p/\hbar} W_2(x-(p/m)\tau_3 ,p),
\end{equation}
where $W_2(x-(p/m)\tau_3 ,p)$ is the Wigner function  (Eq.~(\ref{wigner})) after step 2, where we  replaced all $x$ with $x-(p/m) \tau_3$. Note that in order for complete position space mapping, we again require $\omega_2^2 \tau_2 \approx \tau_1^{-1}+\tau_3^{-1}$.
The Gaussian integral over $p$ can be solved analytically and the remaining $\theta$ integration can be solved numerically for $x' \neq 0$.
For $x'=0$, the final position probability distribution can be written as
\begin{equation}\label{rho31}
\langle x | \rho(\tau_3) | x \rangle = \frac{1}{2 \pi \Delta x^2(\tau_3) \sigma_\mathrm{\Lambda2}(\tau_3) \sigma_c(\tau_3)} \left|\mathrm{Ai}\left( \frac{x}{\Delta x(\tau_3)}\right) \ast \exp \left(- \frac{x^2}{4 \sigma_c^2(\tau_3)} \right) \right|^2 \ast \exp \left(- \frac{x^2}{2\sigma_\mathrm{\Lambda}^2(\tau_3)} \right),
\end{equation}
with
\begin{equation}
    \Delta x(\tau_3) = \hbar \left(\frac{k m}{\hbar} \tan(2 \phi_2) \omega_2^2 \tau_2 \right)^{1/3} \frac{\tau_3}{m},
\end{equation}
\begin{equation}
    \sigma_c(\tau_3) = \frac{\hbar}{2 \sigma_x(\tau_1)} \frac{\tau_3}{m},
\end{equation}
and
\begin{equation}
    \sigma_\Lambda(\tau_3) = \sqrt{\sigma_2^2 + \frac{\hbar^2}{\sigma_x^2(\tau_1)} (\bar{n} + \bar{n}^2)} \frac{\tau_3}{m}.
\end{equation}
Fig.~\ref{Figsup6} shows $g_1(x,x_\mathrm{max2},\tau_3)$ (black), with $x_\mathrm{max2}$ being the position of the second largest peak, as a function of an arbitrary position $x$ revealing the oscillating decay of coherence with distance from the second largest peak following the shape of the interference pattern at $\tau_3$ (red). 
Remarkably, the interference peaks maintain strong coherence between each other, while coherence between minima and maxima almost vanishes. 
For Fig.~4 in the main text, we maximize $\Delta x(\tau_3)$ while requiring  $g_1(x_\mathrm{max1},x_\mathrm{max2},\tau_3) > 0.95$ and $N \approx 1.2 \cdot 10^4$ experimental runs to confirm the wave-nature of the particle with 5$\sigma$ confidence.

\bibliographystyle{apsrev4-2}
\bibliography{supplemental}
